\documentclass{article}
\usepackage[latin1]{inputenc}
\usepackage{subfigure}
\usepackage{graphicx}
\usepackage{astats}

\usepackage{graphics}
\usepackage{dsfont}
\usepackage{pifont}
\usepackage{amsfonts, amsmath, amssymb}
\usepackage{enumerate}
\usepackage{stmaryrd}

\newcommand{\BIC}{\mbox{BIC}}
\newcommand{\ICL}{\mbox{ICL}}
\newcommand{\pen}{\mbox{pen}}
\newcommand{\dd}{\mbox{d}}
\newcommand{\Hyp}{{\bf (H)}}
\newcommand{\Esp}{\mathbb{E}}

\newcommand{\Bcal}{\mathcal{B}}
\newcommand{\Hcal}{\mathcal{H}}
\newcommand{\Gam}{\mathcal{G}\text{am}}
\newcommand{\Mcal}{\mathcal{M}}
\newcommand{\Ncal}{\mathcal{N}}
\newcommand{\Pcal}{\mathcal{P}}
\newcommand{\Scal}{\mathcal{S}}

\newtheorem{theo}{Theorem}[section]    
\newtheorem{lem}[theo]{Lemma}

\newtheorem{prop}[theo]{Proposition}

\title{Exact posterior distributions over the segmentation space and
  model selection for multiple change-point detection problems}

\author{G. Rigaill$^{1, 2, 3}$, E. Lebarbier$^{1, 2}$, S. Robin$^{2,
    1}$ \\
  ~\\
  {\small ($^1$) AgroParisTech, UMR 518, F-75005, Paris, {\sc France}} \\
  {\small ($^2$) INRA, UMR 518, F-75005, Paris, {\sc France}} \\
  {\small ($^3$) Institut Curie, Département de Transfert, F-75005 Paris, France}
  }

\date{}

\begin{document}

\maketitle


\begin{abstract}
  In segmentation problems, inference on change-point position and
  model selection are two difficult issues due to the discrete nature
  of change-points. In a Bayesian context, we derive exact,
  non-asymptotic, explicit and tractable formulae for the posterior
  distribution of variables such as the number of change-points or
  their positions. We also derive a new selection criterion that
  accounts for the reliability of the results. All these results are
  based on an efficient strategy to explore the whole segmentation
  space, which is very large. We illustrate our methodology on both
  simulated data and a comparative genomic hybridisation profile.
\end{abstract}

\paragraph{Keywords:} BIC, change-point detection, ICL,  model
selection, posterior distribution of change-points

\paragraph{Short title:} Posterior distribution over the segmentation space

\section{Introduction} \label{Sec:Intro}

Segmentation and change-point detection problems arise in many
scientific domains such as econometrics, climatology, agronomy or
molecular biology. The general problem can be written as follows. It
is assumed that the observed data $\{y_t\}_{t=1,\ldots,n}$ is a
realization of an independent random process
$Y=\{Y_t\}_{t=1,\ldots,n}$. This process is drawn from a probability
distribution $G$, which depends on a set of parameters denoted by
$\theta$.  These parameters are assumed to be affected by $K-1$ abrupt
changes, called change-points, at some unknown positions
$\tau_2,\ldots, \tau_K$(with the convention $\tau_1=1$ and
$\tau_{K+1}=n+1$).  Thus, the change-points delimit a partition $m$ of
$\{1,\ldots,n\}$, called here a segmentation, into $K$ segments
$r^{(k)}$ such that $r^{(k)}=\llbracket \tau_k, \tau_{k+1} \llbracket
= \{\tau_k, \tau_k+1, \dots, \tau_{k+1}-1\}$ and
$$
m = \{r^{(k)}\}_{k=1,\ldots,K}
$$
The segmentation model has the following general form for a given $m$:
$$
Y_t \sim G(\theta_{r})  \qquad \mbox{if } t \ \in \ r \quad \text{and}
\quad r \in m
$$
where $\theta_r$ stands for the parameters of segment $r$.
In this study, all the change-points are detected simultaneously, a
strategy called off-line detection (as opposed to on-line detection).
With this strategy, the question of finding the best segmentation in a
given number of segments has already been largely studied (see for
example \cite{Lav05}, \cite{BBM00}, \cite{BP03}).
But two important issues remain: assessing the quality of the proposed
segmentation and selecting the number of segments (also called
dimension). In both cases, the main problem is the discrete nature of
the change-points, which prevents the use of routine statistical
inference.

On the one hand, the quality of a given segmentation can be assessed
by studying the uncertainty of the change-point positions. From a
non-asymptotic and non-parametric point of view, the standard
likelihood-based inference is very intricate, since the required
regularity conditions for the change-point parameters are not
satisfied (\cite{Feder75}). Different methods to obtain change-point
confidence intervals have been proposed. Most of them are based on the
limit distribution of the change-point estimators (\cite{Feder75},
\cite{BP03}) or the asymptotic use of a likelihood-ratio statistic
(\cite{Mug03}). Others proposed confidence intervals are based on
bootstrap techniques (\cite{HusHir08} and references therein). A
practical comparison of these methods can be found in
\cite{TomLesp03}.

On the other hand, choosing the number of segments is also a
critical issue.  This is usually done by minimising a penalised
contrast function and the problem is to find a good penalty. General
penalized criteria have been developed, such as AIC
(\cite{Akaike73}) and BIC (\cite{Sch78}). In the segmentation
framework, these criteria are not adapted since an exponential model
collection is considered (\cite{BM04}, \cite{BGH09}) and these
criteria tend to overestimate the number of segments (see for
example \cite{Lav05}). Recently, some penalised criteria have been
proposed specially for the segmentation framework. Some depend on
constants to be calibrated (\cite{Lav05} and \cite{Leb05}), but
others do not (\cite{ZhS07}).  More precisely, \cite{ZhS07}
discussed the fact that the classical BIC was not theoretically
justified in the segmentation context. Indeed, the BIC criterion is
derived from an asymptotic approximation of the posterior model
probabilities and requires the likelihood function to be three times
differentiable with respect to the parameters of the model
(\cite{Kass95}, \cite{LeM06}).  As the change-points are discrete
parameters, the previous condition is not satisfied.  A modified BIC
criterion has thus been developed by \cite{ZhS07} by considering a
continuous-time version of the problem.

The purpose of our work is to provide exact, non-asymptotic, explicit
and tractable formulae for both the posterior probability of a
segmentation and that of a change-point occurring at a given position.
More specifically, we consider the segmentation problem in a Bayesian
framework so that the posterior probability of a segmentation is well
defined. To tackle the discrete nature of change-points, we work at
the segment level, where statistical inference is straightforward.
From these segments, the issue is to get back to the segmentation or
dimension level. Provided that the segments are independent, it will
be necessary to calculate quantities such as:
\begin{equation} \label{Eq:eq_base}
\sum_{m \in \Mcal^{\star}} P(Y|m) P(m)
 =
\sum_{m \in \Mcal^{\star}} P(m) \prod_{r \in m} P(Y^r|r)
\end{equation}
where $Y^r$ stands for all observations in segment $r$ and
$\Mcal^{\star}$ is usually a very large set of segmentations. We
propose a close-form (in terms of matrix products) and tractable
formulation of such quantities.  Some similar quantities were computed
by \cite{Gue08} in a non-Bayesian context, using a
forward-backward-like algorithm. However, this author computes all
these quantities for fixed values of the segment parameters, which are
the maximum likelihood estimators. From our formula, we derive key
quantities to assess the quality of a segmentation and select the
number of segments.

On the one hand, we obtain the exact formulae for both the posterior
probability of a segmentation and that of a change-point occurring at
a given position. This enables the construction of credibility
intervals for change-points.  Moreover, we retrieve the exact
posterior probability of a segment within a given dimension, the exact
entropy of the posterior distribution of the segmentations within a
given dimension and the exact posterior mean of the signal.

On the other hand, we derive a so-called 'exact' BIC criterion for
choosing the number of segments $K$, taking $\Mcal^{\star}=\Mcal_K$
which is the set of all possible segmentations with $K$ segments.  In
the same way, we derive the ICL criterion of \cite{BCG00} in the
segmentation framework. This last criterion takes into account the
reliability of the results.

In Section \ref{Sec:ProbaSeg}, we give some exact formulae to explore
the segmentation space and assess the quality of a segmentation. In
Section \ref{Sec:Select}, we focus on the model selection problem: we
derive an exact BIC criterion and propose a new ICL criterion. In the
last section, we illustrate our results first on Poisson simulated
data and second on comparative genomic hybridization (CGH) data in a
Gaussian framework.

\section{Exploring the segmentation space}
\label{Sec:ProbaSeg}

A naive computation of \eqref{Eq:eq_base} is impossible when
$\Mcal^{\star}$ is large, which is usually the case.  For example, if
$\Mcal^{\star} = \Mcal_K$, there are $\binom{n-1}{K-1}$ segmentations
of $n$ data into $K$ segments.  In this section we propose a tractable
and close-form formula of \eqref{Eq:eq_base}.  The following
assumption enables us to derive an exact matrix product formulation of
\eqref{Eq:eq_base} enabling its straightforward computation in $O(K
n^2)$ time.

\paragraph{Factorability assumption:} A model satisfies the
factorability assumption if
\begin{equation} \label{Eq:Factor}
\Hyp: P(Y, m) = C \prod_{r \in m} a_r P(Y^r|r)
\end{equation}
where $P(Y^r|r) = \int P(Y^r | \theta_r) P(\theta_r) \dd \theta_r$. In
the following, for the sake of clarity, we will simply denote
$P(Y^r)$.  This is true when all segment parameters are different but
this is false, for example, for the normal homoscedastic model
$G(\theta_r) = \Ncal(\mu_r, 1/\tau)$ with unknown precision $\tau$.

We denote by $\Mcal_K(\llbracket i, j \llbracket)$ the set of all
possible segmentations of $\llbracket i, j \llbracket$ into $K$
segments. The simplified notation $\Mcal_K$ refers to
$\Mcal_K(\llbracket 1, n+1 \llbracket)$.

\begin{theo} \label{Thm:Complexity}
  Consider a function $F$ such that, for all $k \in \llbracket 1, K
  \rrbracket $ and for all segmentation $ m \in \Mcal_k(\llbracket 1,
  j\llbracket)$ (for $1 \leq j \leq n+1$), there exists a function $f$
  such that: $ F(m) = \prod_{r \in m} f(r)$. Let $\mathbf{A}$ be a
  square matrix with $n+1$ columns such that
  $$
  \begin{array}{rclll}
  {\bf A}_{ij} & = & f(\llbracket i, j \llbracket ) & \quad & \text{if } 1
  \leq i < j \leq n+1 \\
  & = & 0 & & \text{otherwise.}
  \end{array}
  $$
  Then,
  $$
  \sum_{m \in \Mcal_k(\llbracket 1, j\llbracket)} F(m) = (\mathbf{A}^k)_{1,j}
  $$
  and the $K \times (n+1)$ elements of
  $$
  \left\{ \sum_{m \in \Mcal_k(\llbracket 1, j\llbracket)} F(m)
  \right\}_{k \ \in \ \llbracket 1, K \rrbracket \ \cap \ j \ \in \
    \llbracket 1, n+1 \rrbracket }
  $$
  can all be computed in $O(K n^2)$
\end{theo}

The proof is given in Appendix \ref{App:ProofComplexity}. It is based
on a linear algebra lemma. The lower triangular part of matrix ${\bf
  A}$ is set to 0 to fit the segmentation context.  Note that,
similarly, we have $ \sum_{m \in \Mcal_k(\llbracket i, j\llbracket)}
F(m) = (\mathbf{A}^k)_{i,j}$ for all $1 \leq i \leq j \leq n+1$.
Theorem \ref{Thm:Complexity} will be used many times in the following
sections, using a specific function $f(r)$ for each quantity of
interest.

\subsection{Joint distribution of the data and the segmentation or the
  dimension}

$P(Y, m)$ and $P(Y, K)$ are key ingredients to calculate various
quantities of interest, such as \eqref{Eq:eq_base}. To calculate $P(Y,
m)$ and
\begin{equation} \label{Eq:pYK}
P(Y, K) = \sum_{m \in \Mcal_K} P(Y, m),
\end{equation}
we first need to define {priors} for the segmentation $m$. We consider
here two typical {priors}.
\begin{description}
\item[Uniform conditional on the dimension:] For any {prior} on the
  dimension $P(K)$, we define a uniform prior distribution for $m$
  given its dimension $K$:
  \begin{equation} \label{Eq:PriorUnif|K}
    P(m|K(m)) = \binom{n-1}{K(m)-1}^{-1} \qquad \Rightarrow \qquad P(m) = P(K(m))
    \left/ \binom{n-1}{K(m)-1} \right.
  \end{equation}
  that is $a_r = 1$ in \eqref{Eq:Factor}, denoting $K(m)$ the number
  of segments (i.. the dimension) of the segmentation $m$.
\item[Homogeneous segment lengths:] Segmentation with balanced segment
  lengths are sometimes desirable. They are favoured by the following
  {prior}:
  \begin{equation} \label{Eq:PriorEntrop}
    P(m) = C \prod_{r \in m} n_r^{-1}, \qquad
    \text{where $C$ ensures that} \sum_{m \in \Mcal} P(m) = 1.
  \end{equation}
  that is $a_r = n_r^{-1}$ in \eqref{Eq:Factor}, where $n_r$ denotes
  the length of segment $r$ and $\Mcal$ the set of all considered
  segmentations.  In this case, the prior distribution of $m$ is
  directly defined and the prior distribution of the dimension $P(K)$
  is not explicit.  Determining the constant $C$ requires summing over
  all possible segmentations.  This sum can be handled using the
  properties given below.
\end{description}

\begin{prop} \label{Prop:pYm-pYK}
  Under assumption $\Hyp$, for prior distributions
  \eqref{Eq:PriorUnif|K} and \eqref{Eq:PriorEntrop}, $P(Y, K)$ can be
  computed in $O(K n^2)$ as $P(Y, K) = C (\mathbf{A}^k)_{1, n+1}$ with
  ${\bf A}_{i,j} = 0$ for $j \leq i$ and, for $j > i$, for prior
  distribution \eqref{Eq:PriorUnif|K}:
  $$
  \mathbf{A}_{i,j} = P(Y^{\llbracket i, j \llbracket}) \quad \mbox{and} \quad C^{-1}  =  \binom{n-1}{K-1};
  $$
  and for prior distribution \eqref{Eq:PriorEntrop}:
  $$
  \mathbf{A}_{i,j} = n_{\llbracket i, j \llbracket}^{-1}
  P(Y^{\llbracket i, j \llbracket}) \quad \mbox{and} \quad
  C^{-1} = \sum_{m \in \Mcal_K} \prod_{r \in m} n_r^{-1}.
  $$
\end{prop}


\paragraph{Proof.}
For prior distribution \eqref{Eq:PriorUnif|K}, we use Theorem
\ref{Thm:Complexity} with $f(r) = P(Y^r)$, implying $\mathbf{A}_{i,j}
= f(\llbracket i, j \llbracket ) = P(Y^{\llbracket
  i, j \llbracket})$.  \\
For prior distribution \eqref{Eq:PriorEntrop}, we first retrieve $C$
using Theorem \ref{Thm:Complexity} with $f(r) = n_r$. The result
follows, using Theorem \ref{Thm:Complexity} again, taking $f(r) =
n_r^{-1} P(Y^r)$.$\blacksquare$


\bigskip\bigskip
The preceding results require the calculation of $P(Y^r)$. Hence,
$n(n-1)/2$ integrals need to be evaluated, corresponding to each
possible segment. For general {priors}, they can be evaluated
numerically or via any stochastic algorithm. A close form can be
obtained if conjugate {priors} are used.

\paragraph{Poisson and Gaussian models.}\label{PoisGauss}
We recall classical results for two models that will be used later.
First is the segmentation problem of a piecewise constant Poisson rate
model:
\begin{eqnarray}  \label{Eq:ModelPoisson}
  \{\mu_r\} \text{ i.i.d.}, \quad \mu_r & \sim & \Gam(\alpha_r,\beta_r);
  \nonumber \\
  \{Y_t\} \text{ independent}, \quad Y_t & \sim & \Pcal(\mu_r) \quad
  \mbox{if } t \ \in \ r.
\end{eqnarray}
Second is the segmentation of a Gaussian signal where both the mean and
the variance are affected by the change-points:
\begin{eqnarray} \label{Eq:ModelGauss}
  \{\tau_r\} \text{ i.i.d.}, \quad \tau_r & \sim & \Gam(\nu_0/2,2/s_0);
  \nonumber \\
  \{\mu_r\} \text{ independent}, \quad \mu_r|\tau_r & \sim &
  \Ncal(\mu_0,(n_0 \tau_r)^{-1}); \nonumber \\
  \{Y_t\} \text{ independent}, \quad Y_t & \sim & \Ncal(\mu_r, 1/\tau_r)
  \qquad   \mbox{if } t \ \in \ r.
\end{eqnarray}
For the Poisson model, we get
$$
P(Y^r)=\frac{\Gamma(\alpha+\sum_{t \in r} Y^r_t) \
  \beta_r^{\alpha_r}}{(\beta_r+n_r)^{\alpha_r+\sum_{t \in r} Y^r_t} \
  \Gamma(\alpha_r) \prod_{t \in r} (Y^r_t !)}.
$$
For the Gaussian heteroscedastic model, we get
\begin{equation} \label{Eq:PosteriorGauss}
P(Y^r)= \frac{{n_0}^{1/2} \ (s_0/2)^{\nu_0/2} \
\Gamma((\nu_0+n_r)/2)}{(2 \pi)^{n_r/2} \ \Gamma(\nu_0/2) \
\sqrt{n_r+n_0}} \ \theta^{(\nu_0+n_r)/2}
\end{equation}
where $\theta = 2 (n_r S_r^2 + s_0 + \frac{n_r n_0 (\bar{y}_r -
  \mu_0)^2}{n_r + n_0})^{-1}$, $S_r^2=\sum_{t \in r}
(Y_t-\bar{y}_r)^2/n_r$ and $\bar{y}_r$ is the empirical mean of the
signal within segment $r$.


\subsection{Posterior distribution of the change-points and segments }

We now give explicit formulae for the posterior distribution of
change-points and segments. We first define the corresponding
segmentation subsets:
\begin{description}
\item[$\Bcal_{K,k}(t)$] is the subset of segmentations from $\Mcal_K$ such
  that the $k$-th segment starts at position $t$, i.e. that the
  $(k-1)$-th change-point is at $t$:
  $$
  \Bcal_{K,k}(t) = \{m \in \Mcal_K: \tau_k=t\};
  $$
\item[$\Bcal_K(t)$] is the subset of segmentations having a
  change-point at position $t$:
  $$
  \Bcal_K(t) = \bigcup_k \Bcal_{K,k}(t);
  $$
\item[$\Scal_{K,k}(\llbracket t_1, t_2 \llbracket)$] is the subset of
  segmentations having segment $r = \llbracket t_1, t_2\llbracket$ as
  their $k$-th segment:
  $$
  \Scal_{K,k}(\llbracket t_1, t_2 \llbracket) = \{m \in
  \Mcal_K(\llbracket 1, n+1 \llbracket): \tau_k=t_1, \tau_{k+1}=t_2\};
  $$
\item[$\Scal_K(\llbracket t_1, t_2 \llbracket)$] is the subset of
  segmentations including segment $\llbracket t_1, t_2 \llbracket$:
  $$
  \Scal_K(\llbracket t_1, t_2 \llbracket) = \bigcup_k
  \Scal_{K,k}(\llbracket t_1, t_2 \llbracket).
  $$
\end{description}
We denote the conditional probability given the data $Y$ and the
dimension $K$ of each of these subsets by the corresponding capital
letters with same indices, e.g.
$$
B_{K,k}(t) = \Pr\{m \in \Bcal_{K, k}(t) | Y, K\}.
$$
$B_{K}(t)$, $S_{K, k}(t)$ and $S_{K}(t)$ are defined similarly.
The following proposition gives explicit formulae for these
probabilities.

\begin{prop}\label{Prop:IntQuant}
  For all $\llbracket t_1, t_2 \llbracket$ such that $t_1 < t_2$, we
  define, for $K \geq 1$,
  $$
  F_{t_1, t_2}(K) = \sum_{m \in \Mcal_K(\llbracket t_1,
    t_2\llbracket)}P(Y^{\llbracket t_1, t_2 \llbracket} | m) P(m | K),
  $$
  and we set $F_{t_1, t_2}(K) = 0$ if $t_1 \geq t_2$.  Under
  assumption $\Hyp$, the probabilities $B_{K,k}(t)$, $B_{K}(t)$,
  $S_{K, k}(t)$ and $S_{K}(t)$ are
  \begin{eqnarray*}
    B_{K,k}(t) & = & \frac{ F_{1,t}(k-1) F_{t,n+1}(K-k+1) }{P(Y | K)}, \\
    S_{K, k}(t_1, t_2) & = & \frac{F_{1,t_1}(k-1) F_{t_1,t_2}(1)
      F_{t_2,n+1}(K-k)}{P(Y | K)}
  \end{eqnarray*}
  $B_{K}(t) =\sum_{k=1}^K B_{K,k}(t)$ and $S_{K}(t_1, t_2) = \sum_k
  S_{K,k}(t_1, t_2)$.
\end{prop}

The proof is given in Appendix \ref{App:ProofIntQuant}. It is mainly
based on set decompositions, such as
\begin{equation} \label{Eq:DecompBKk}
\Bcal_{K, k}(t) = \Mcal_{k-1}(\llbracket 1, t\llbracket) \times
\Mcal_{K-k+1}(\llbracket t, n+1 \llbracket)
\end{equation}
and all sums over $\Mcal_{k-1}(\llbracket 1, t\llbracket)$ and
$\Mcal_{K-k+1}(\llbracket t, n+1 \llbracket)$ can be obtained with
Theorem \ref{Thm:Complexity}.

$\{B_{K, k}(t)\}_t$ provides the exact posterior distribution of the
starting point of the $k$-th segment, given dimension $K$. From
that, we get the exact credibility of interval $\llbracket t_1, t_2
\rrbracket$ for change-point $\tau_k$:
$$
C_{K, k}(\llbracket t_1, t_2 \rrbracket) = \Pr\{\tau_k \in
\llbracket t_1, t_2 \rrbracket | Y, K\} = \sum_{t=t_1}^{t_2} B_{K, k}(t).
$$

\subsection{Retrieving the mean signal}
In many applications, the mean value $\mu_t$ of the signal at a given position
can also provide some insight about the phenomenon under study. This
mean signal can be retrieved via model averaging over the segmentation
space. The posterior mean of the signal is
\begin{equation} \label{Eq:posterior_mean}
\bar{s}_K(t) = \sum_{m \in \Mcal_K} P(m | Y, K) \widehat{s}_m(t),
\end{equation} where $\widehat{s}_m(t) = \Esp[\mu_t | m, Y]$.

\begin{prop}
  The posterior mean of the signal given the dimension is
  $$
  \bar{s}_K(t) = \sum_{r \ni t} S_K(r) \widehat{\mu}_r,
  $$
  where $\widehat{\mu}_r = \Esp[\mu_r | Y^r]$. Under assumption
  $\Hyp$, it can be computed with a quadratic complexity.
\end{prop}

\paragraph{Proof.} If a segment $r$ belongs to a
segmentation $m$ and if position $t$ lies in segment $r$ then
$\widehat{s}_m(t) = \widehat{\mu}_r$.  The rest of the formula is
straightforward.  Assumption $\Hyp$ ensures that the $S_K(r)$ can be
computed in $O(K n^2)$.  $\blacksquare$

\subsection{Posterior entropy}

Segmentation problems are often reduced to choosing $\widehat{m}_K$,
the best segmentation (i.e. the one with maximal posterior probability) with
dimension $K$. The other segmentations with dimension $K$ are rarely
considered.  The entropy of the distribution $P(m|Y, K)$
$$
\Hcal(K) = - \sum_{m \in \Mcal_K} P(m | Y, K) \log P(m | Y, K)
$$
measures how the posterior distribution is concentrated around the
best segmentation. Intuitively, a small entropy $\Hcal(K)$ means that
the best segmentation is a much better fit to the data than any other
segmentation. We use this information in Section \ref{Sec:Select} for
model selection.

\begin{prop}
  Under assumption $\Hyp$, the posterior entropy $\Hcal(K)$ is
  $$
  \Hcal(K) = -\sum_{r } S_K(r) \log f(r) + \log A_K
  $$
  where $f(r) = a_r P(Y^r)$ and $A_K = \sum_{m \in \Mcal_K}
  \prod_{r \in m} f(r)$, which can be computed using Proposition
  \ref{Prop:pYm-pYK}.
\end{prop}

\paragraph{Proof.}
Since all distributions can be factorized, we have
\begin{eqnarray*}
  \Hcal(K) & = & -\sum_{m \in \Mcal_K} \sum_{r \in m} P(m | Y,
  K)\log f(r) + \sum_{m \in \Mcal_K} P(m | Y, K) \log A_K
  \\
  & = & -\sum_{r} \log f(r) \sum_{m \in \Mcal_K, m \ni r} P(m | Y, K) +
  \log A_K \sum_{m \in \Mcal_K} P(m | Y, K)
\end{eqnarray*}
and the result follows.  $\blacksquare$

\section{Model selection}
\label{Sec:Select}

In a Bayesian framework, the BIC criterion aims to choose the model
which maximises $P(M|Y)$, where $M$ is the model.  To calculate the
BIC criterion, one needs to know $P(Y |M) = \int P(Y | \theta_M, M)
P(\theta_M | M) \dd \theta_M$, where $\theta_M$ is the set of
parameters of the model $M$. Similar quantities are involved in the
Bayes factor for model comparison (\cite{Kass95}).

In our case, the word 'model' is too broad and we have to distinguish
between the selection of the dimension $K$ and the selection of the
segmentation $m$. When considering the choice of $K$, a direct
application of the Laplace approximation is not theoretically
justified to calculate the previous integral because the required
differentiability condition is not satisfied for change-points
(\cite{ZhS07}).  However, we can bypass the problem by working at the
segment level and then going back at the dimension level using
Proposition \ref{Prop:pYm-pYK}.  Thus, the derivation of BIC criteria
only requires the calculation of $P(Y^r)= \int P(Y^r | \theta_r)
P(\theta_r) \dd \theta_r$, which can be obtained in a close form for
the Poisson model and the heteroscedastic Gaussian model as shown in
Section \ref{PoisGauss}.  Moreover, we derive an adaptation of the ICL
criterion, first proposed for mixture models, to the segmentation
context (\cite{BCG00}).

\subsection{Exact BIC criterion for dimension and segmentation selection}

\paragraph{Choice of  the dimension.}
In segmentation problems, the selection of the 'best' number of
segments $K$ can be addressed per se, or as a first step toward the
selection of the 'best' segmentation. The Bayesian framework suggests
to choose
\begin{equation} \label{Eq:BestK}
\widehat{K} = \underset{K}{\arg\min} \; \BIC(K), \quad
  \text{where} \quad \BIC(K) = - \log P(Y, K).
\end{equation}
$\BIC(K)$ can be computed in a quadratic time, using Proposition
\ref{Prop:pYm-pYK}.

\paragraph{Choice of the segmentation.}
The best segmentation can be chosen in two ways.
\begin{description}
\item[{\sl Two-step strategy:}] The 'best' segmentation $m$ can be chosen,
  conditionally to the pre-selected dimension $\widehat{K}$ as
  \begin{equation} \label{Eq:Bestm}
    \widehat{m}(\widehat{K}) = \underset{m \in
    \Mcal_{\widehat{K}}}{\arg\min} \;
    \BIC(m|\widehat{K}), \quad
    \text{where} \quad \BIC(m|\widehat{K}) = - \log P(Y, m |
    \widehat{K}).
  \end{equation}
\item[{\sl One-step strategy:}] The 'best' segmentation $m$ can also be
  directly chosen among a larger collection $\Mcal = \bigcup_{k=1}^K
  \Mcal_k$ as
\begin{equation} \label{Eq:Bestm}
  \widehat{m} = \underset{m \in \Mcal}{\arg\min} \; \BIC(m), \quad
  \text{where} \quad \BIC(m) = - \log P(Y, m).
\end{equation}
\end{description}
Both $\BIC(m|K)$ and $\BIC(m)$ can be computed efficiently thanks to
Proposition \ref{Prop:pYm-pYK}.

\subsection{ICL criterion for dimension selection}

In the framework of incomplete data models (e.g. mixture models),
\cite{BCG00} suggest to use the criterion $\ICL(M)$, which is an
estimate of $\Esp[\log P(Y, Z, M) | Y]$ where $Z$ stands for the
unobserved variables. Based on the equation
$$
\Esp[\log P(Y,Z|M) | Y] = \log P(Y|M) + \Esp[\log P(Z|Y, M) | Y],
$$
they argue that the entropy $H(Z|Y, M) = -\Esp[\log P(Z|Y, M)|Y]$ is an
intrinsic penalty term. The $\ICL$ criterion will tend to select models that provide
a reliable prediction of $Z$, i.e. with a small entropy. This may be
desirable, for example in the classification context.

In the segmentation context, the segmentation $m$ can be considered as
an unobserved variable. The dimension $K$ can then be chosen according
to the $\ICL$ as
$$
\widehat{K} = \arg\min_{K} \ICL(K)
\quad \text{where} \quad
\ICL(K) = - \log P(Y, K) + H(m|Y, K).
$$\cite{BCG00}
We expect $\ICL$ to favour the dimension $K$ where the best
segmentation $\widehat{m}(K)$ clearly outperforms the other
segmentations in $K$ segments, so that $\widehat{m}(K)$ is more
reliable.

\subsection{Comparison with other penalized criteria}

Many model selection criteria have the following form:
$$
\log P(Y | \widehat{\theta}, m) - \pen(m)
$$
and use a two-step strategy. Interestingly, since the penalty generally
depends only on the dimension (\cite{Leb05}, \cite{Lav05}), the best
segmentation $\widehat{m}(K)$ does not actually depend on the penalty.

The calculation of the exact BIC does not provide any explicit penalty
enabling a direct comparison with such criteria. For such comparison, we
derive two approximations of $\log P(Y^r) = \log \int P(Y^r|\theta_r)
P(\theta_r) \dd \theta_r$ in the heteroscedastic Gaussian case. The
first one is based on a Laplace approximation:
$$
\log P(Y^r) \approx \log P(Y^r | \widehat{\theta}_r) -
\frac{D}{2} \log n_r
$$
where $D$ stands for the number of parameters involved in each
segment (here, $D=2$).  This approximation is valid only for large
segments, i.e. where $P(Y^r|\theta_r)$ satisfies regularity
conditions.  For the second approximation, we let the hyperparameters
$n_0, \nu_0$ and $S_0$ go to 0 in \eqref{Eq:PosteriorGauss} and we
obtain
$$
\log P(Y^r) \approx - \frac{n_r}2 \log S^2_r - \frac{D}{2} \log n_r
\approx \log P(Y^r | \widehat{\theta}_r) - \frac{D}{2} \log n_r.
$$
We emphasize that these approximations are both questionable since
the asymptotic framework of the Laplace approximation is not correct
for small segments and because the {priors} are improper for null
hyperparameters. Our purpose is only to show that they both provide
the same penalty form:
$$
\log P(m|Y) \approx \log P(m) + \log P(Y|\widehat{\theta}, m) -
\frac{D}{2}\sum_{r \in m} \log n_r.
$$
Using uniform prior \eqref{Eq:PriorUnif|K}, we get
$$
\pen(m) = \log P(K(m)) -\log \binom{n-1}{K(m)-1} - \frac{D}{2}
\sum_{r
  \in m}\log n_r.
$$
A similar form is obtained in the Poisson case. The complexity
term, $\log \binom{n-1}{K-1}$, is similar to the one of \cite{Leb05}.
The regularity term, $\sum_{r \in m}\log n_r$, favours segments with
equal lengths and is similar to the one of \cite{ZhS07}. Using the
alternative prior \eqref{Eq:PriorEntrop} reinforces the regularity
term. Due to this term, the best segmentation $\widehat{m}(K)$ within
$\Mcal_K$ does depend on the penalty.

\section{Applications}
\label{Sec:Appli}

In this section, we first present a simulation study to assess the
ability of the exact BIC and ICL criteria to select the dimension and
the ability of model averaging to retrieve the mean signal. We then
analyse a real CGH profile and use our formulae to assess the quality
of the segmentation.

\subsection{Simulations}

\paragraph{Simulation design.}
We performed the simulation study in the Poisson model
\eqref{Eq:ModelPoisson} so that only one parameter had to be chosen.
We simulated a sequence of $150$ observations affected by six
change-points at the following positions: $21$, $29$, $68$, $82$,
$115$, $135$. Odd segments had a mean of 1, while even segments had a
mean of $1+\lambda$, where lambda varies from 0 to 10. The higher
$\lambda$ is, the easier it should be to recover the true number of
change-points. The hyperparameters $\alpha$ and $\beta$ were set to be
equal and we considered three values for them: 0.01, 0.1 and 1. For
each configuration, we simulated 300 sequences.





\subsection{Recovering the number of change-points}

\subsubsection{The ICL criterion performed better than the BIC criterion}

\begin{figure}[Ht]
\centering
\includegraphics[scale=0.3]{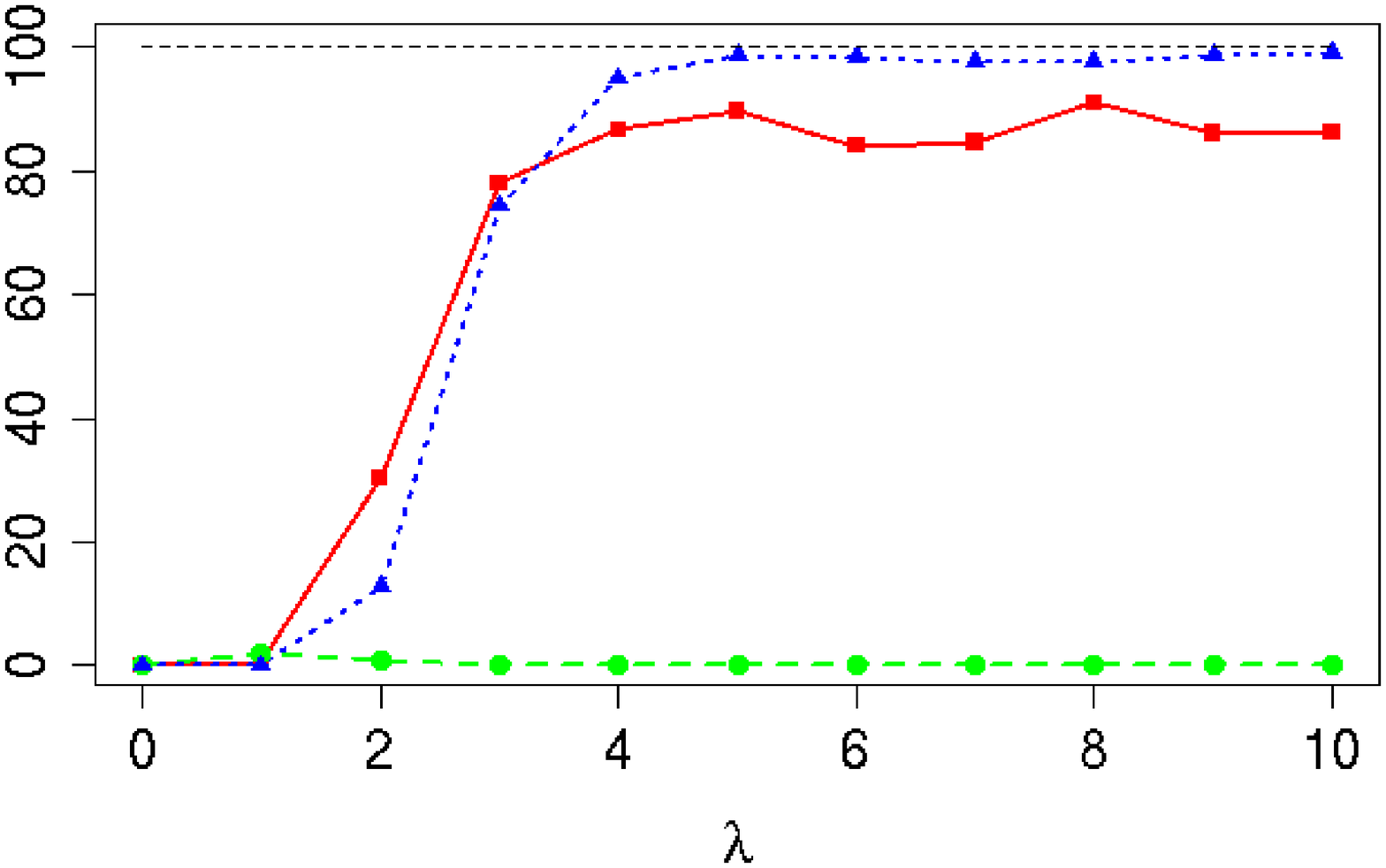}
\includegraphics[scale=0.3]{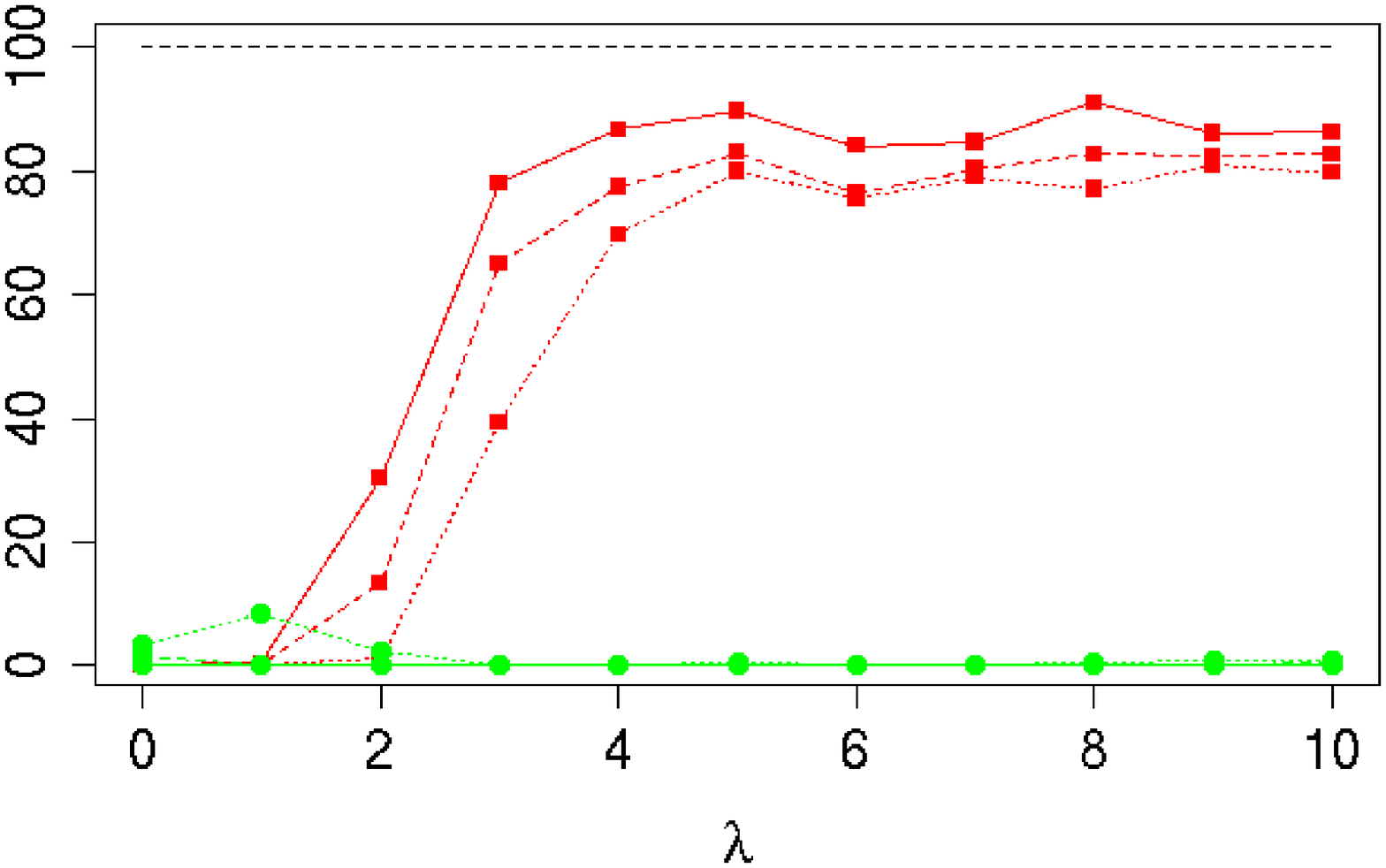}
\caption{Percentage of true dimension recoveries
  as a function of $\lambda$. Left panel: for the three criteria.
  $\BIC(\widehat{m}_K): \blacksquare-$, $\BIC(K): \bullet- -$ and
  $\ICL(K): \blacktriangle\cdots$.  Right panel:
  for the BIC criteria; $\bullet$: uniform prior over all
  segmentations, $\blacksquare$: uniform prior over all segmentations
  of a dimension,  $-: \alpha = \beta = 1$, $--: \alpha = \beta =
  0.1$, $\cdots: \alpha = \beta = 0.01$.\label{Fig:PctRecover}}
\end{figure}

\paragraph{Model selection.}
The BIC criterion for dimension selection, $\BIC(K)$, almost never
returned the true dimension, even for high values of $\lambda$ (Figure
\ref{Fig:PctRecover}, where $\alpha$ and $\beta$ were set to $1$).  On
the other hand, both the BIC criterion for model selection, $\BIC(m)$,
and the ICL criterion, $\ICL(K)$, tend to recover the true dimension
more often when $\lambda$ became larger.  $\ICL(K)$ even increased to
a maximum of $99 \%$ true recoveries compared to a maximum of $91 \%$
for the $\BIC(m)$ criterion for model selection.

\paragraph{Influence of the {priors}.}
The ability of $\BIC(m)$ to retrieve the true dimension was greatly
affected by the prior distribution of the segmentation (Figure
\ref{Fig:PctRecover}).  To illustrate this effect, we considered a
{prior} that gave equal probability to all segmentations, whatever
their dimension: $P(m) = \text{cst}$.  This led to a $90 \%$
decrease in the ability to return the true dimension compared to a
conditional uniform {prior} given the dimension
\eqref{Eq:PriorUnif|K} (with $P(K(m)) = \text{cst}$ whatever $m$).
The impact of the two hyperparameters $\alpha$ and $\beta$ seemed
relatively limited in comparison: less than $10 \%$ difference in
the ability to return the true dimension (Figure
\ref{Fig:PctRecover}).

\paragraph{Estimation of the mean signal.}
We then compared the ability of the maximum likelihood estimators
(MLE) and that of the posterior mean signal to recover the true signal
in terms of the Kullback-Leibler distance. For each simulation, we
computed the following:
$$
d(\widehat{\mu}, \mu) = \sum_t KL[\Pcal(\widehat{\mu}_t); \Pcal(\mu_t)]
$$
for both the MLE estimate $\widehat{\mu} = \widehat{\mu}_\text{MLE}$
and the posterior mean $\widehat{\mu} = \bar{s}_K(t)$ (see equation
(\ref{Eq:posterior_mean})).

\begin{figure}[Ht]
  \centering \includegraphics[scale=0.3]{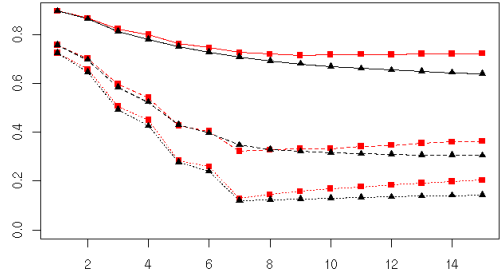}
  \caption{Kullblack-Leibler-based distance $d$ to the true signal as
    a function of the dimension. $\blacksquare$:
    $d(\widehat{\mu}_\text{MLE}, \mu)$, $\blacktriangle$:
    $d(\overline{\mu}, \mu)$ for three value of $\lambda$ $1$: $-$,
    $2$: $--$ and $6$: $\cdots$. The true number of segments was
    7.\label{Fig:Dist}}
\end{figure}

When $K$ was lower than the true dimension ($7$ segments), the two
estimates were almost equivalent (Figure \ref{Fig:Dist}). However,
for larger dimensions, the distance of the MLE to the true signal
increased whereas the distance of the posterior mean did not (Figure
\ref{Fig:Dist}).  The posterior mean seemed less prone to
over-fitting.  Moreover, for a very small signal-to-noise ratio
($\lambda=1$), the distance between the posterior mean of the signal
and the true signal still decreased when $K$ was higher than the
true dimension.  Therefore, when the signal was of poor quality and
led to a poor assessment of the true dimension, the posterior mean
of the signal led to better results. Moreover, the standard
deviation of $d$ for the posterior mean is almost always smaller
than the one of the MLE (not shown).

\subsection{Analysis of a CGH profile}
\begin{figure}[Ht]
\centering
\includegraphics[scale=0.22]{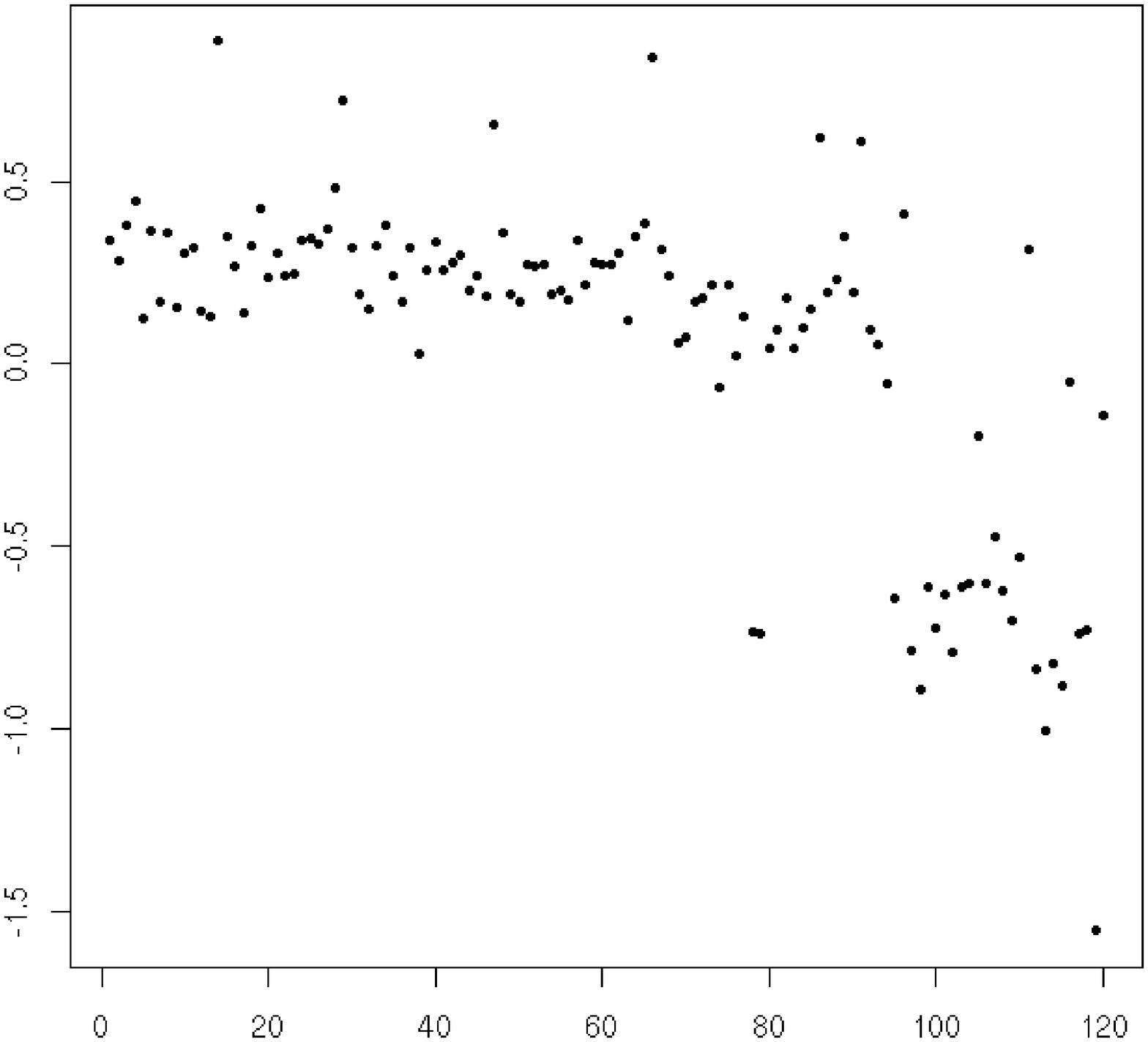}
\centering \includegraphics[scale=0.22]{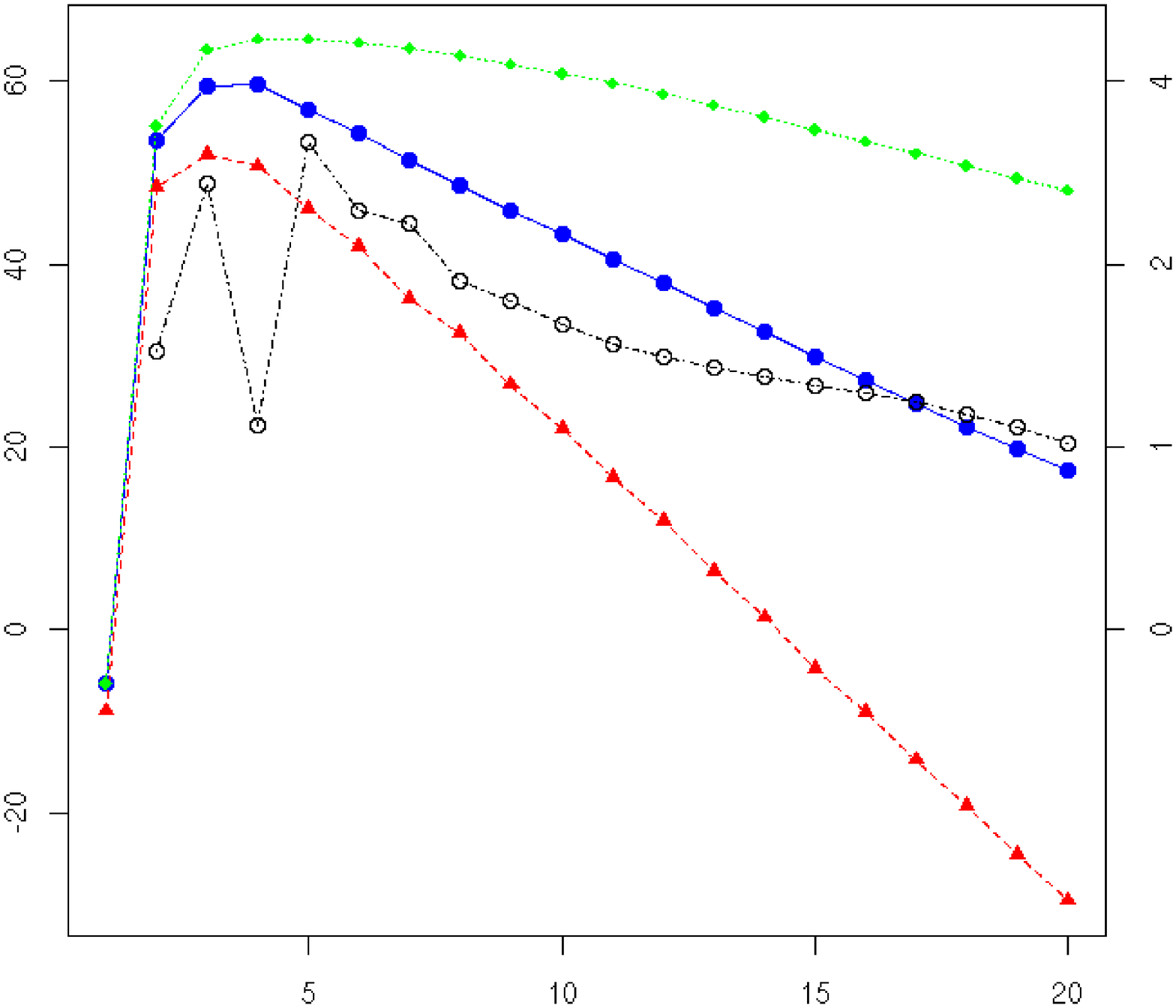}
\caption{Left panel: Chromosome 10 profile of cell line BT474.
  The DNA copy number logratio is represented as a function of its
  position along the chromosome. Right panel: (Left axis) $\BIC(m)$:
  $\blacktriangle$, $\BIC(K)$: $\bullet$ and $\ICL(K)$: $\blacksquare$
  as a function of the dimension. (Right axis) $\Hcal(K) -
  \Hcal(K-1)$: $\circ$ as a function of the dimension.
\label{Fig:ProfilExemple}}
\end{figure}

In the following subsection, we used a comparative genomic hybridation
(CGH) profile to illustrate our methodology.  CGH enables the study of
DNA copy number gains and losses along the genome (\cite{PSS98}). We
used the Gaussian segmentation model defined in \eqref{Eq:ModelGauss}
that is often used for this type of data (\cite{PRL05}). The profile
shown in Figure \ref{Fig:ProfilExemple} represents the copy
number logratio of cell line BT474 to a normal
reference sample, along chromosome 10.

\begin{figure}[Ht]
  \centering
  \begin{tabular}{ccc}
    $(i)$ &
    \begin{tabular}{c}
      \includegraphics[height=3.4cm, width=6cm]{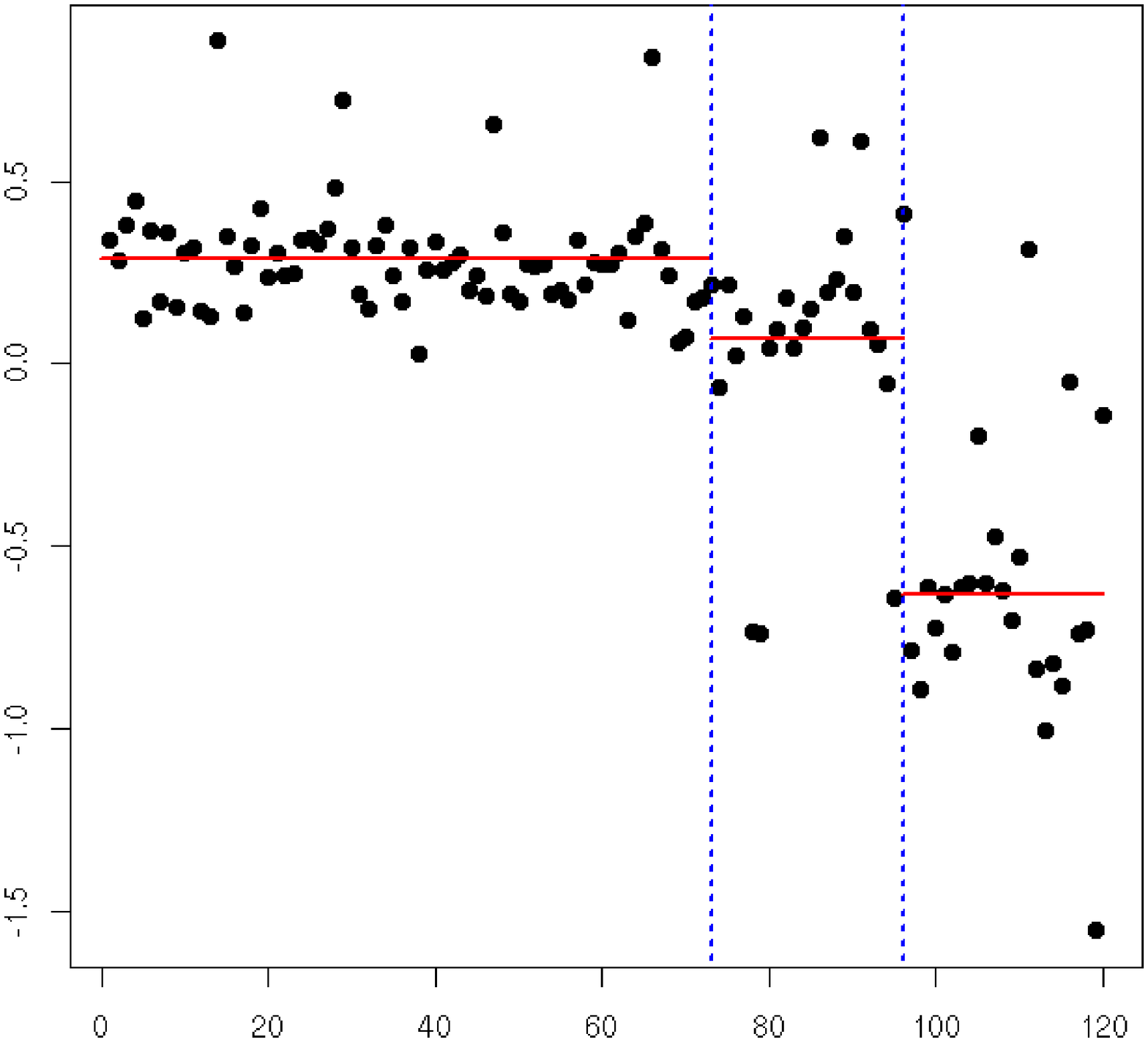}
    \end{tabular}
    &
    \begin{tabular}{c}
      \includegraphics[height=3.4cm, width=6cm]{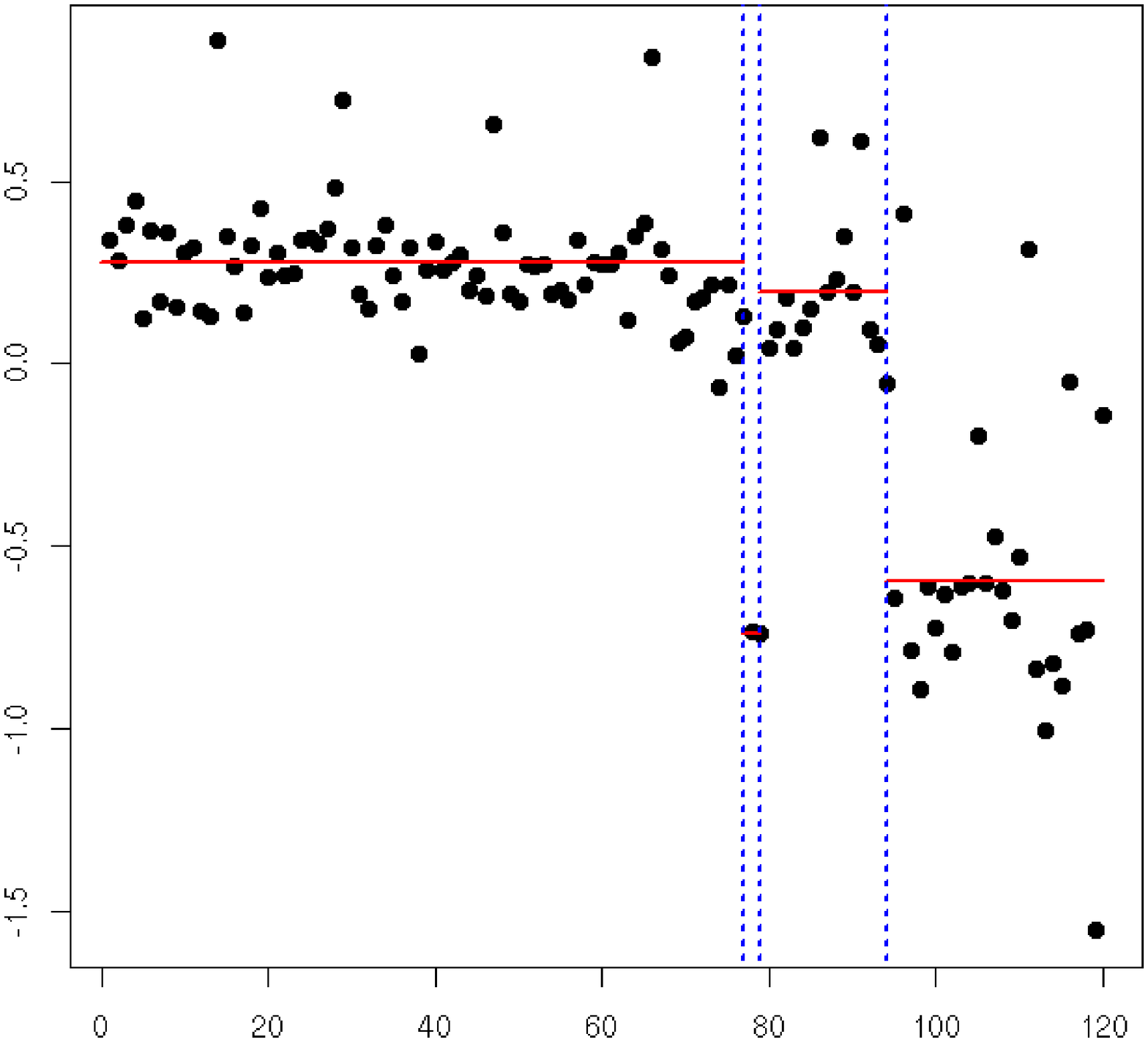}
    \end{tabular}
    \\
    $(ii)$ &
    \begin{tabular}{c}
      \includegraphics[height=3.4cm, width=6cm]{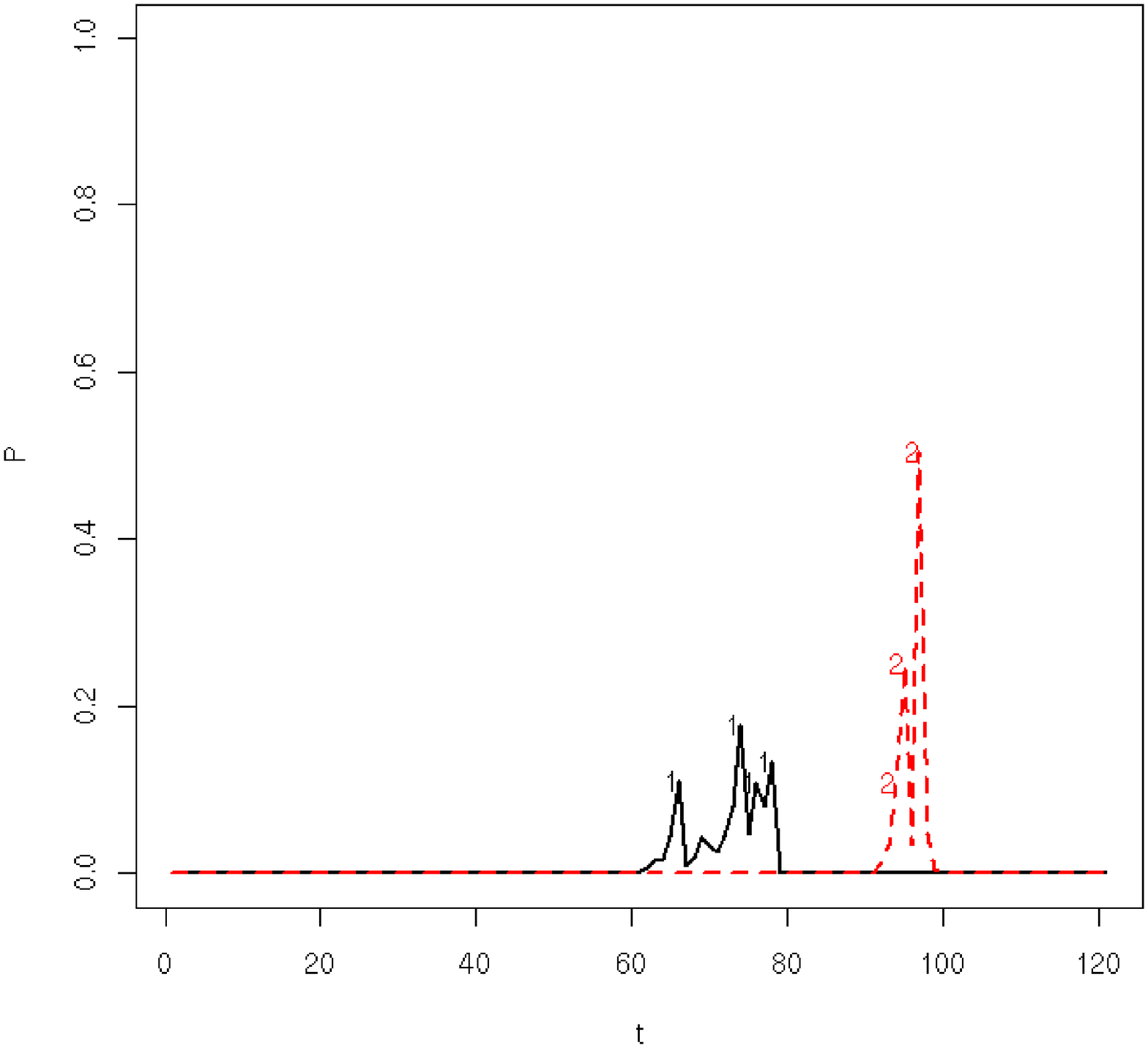}
    \end{tabular}
    &
    \begin{tabular}{c}
      \includegraphics[height=3.4cm, width=6cm]{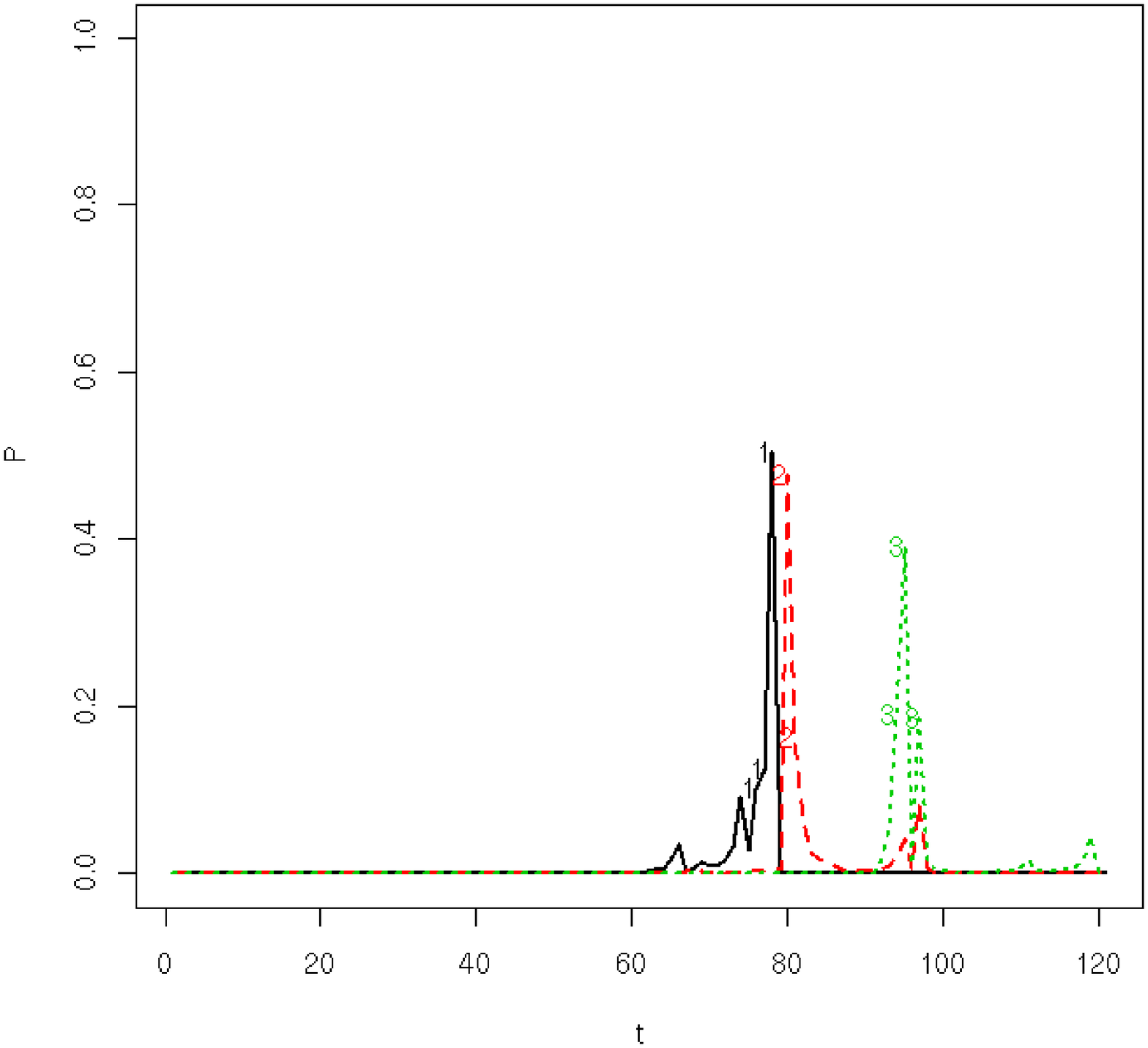}
    \end{tabular}
    \\
     $(iii)$ &
     \begin{tabular}{c}
       \includegraphics[height=3.4cm,
       width=6cm]{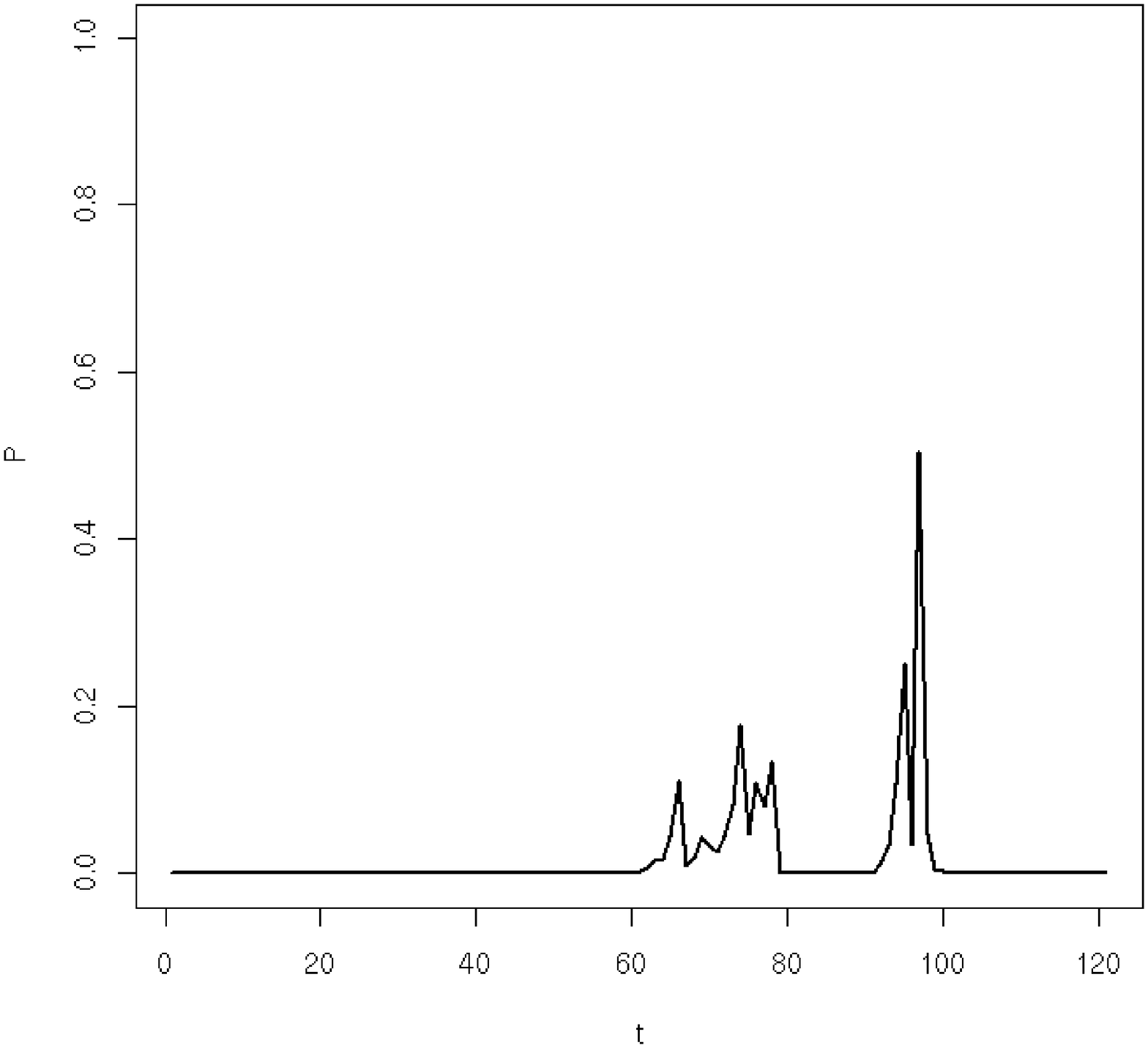}
     \end{tabular}
     &
     \begin{tabular}{c}
       \includegraphics[height=3.4cm,
       width=6cm]{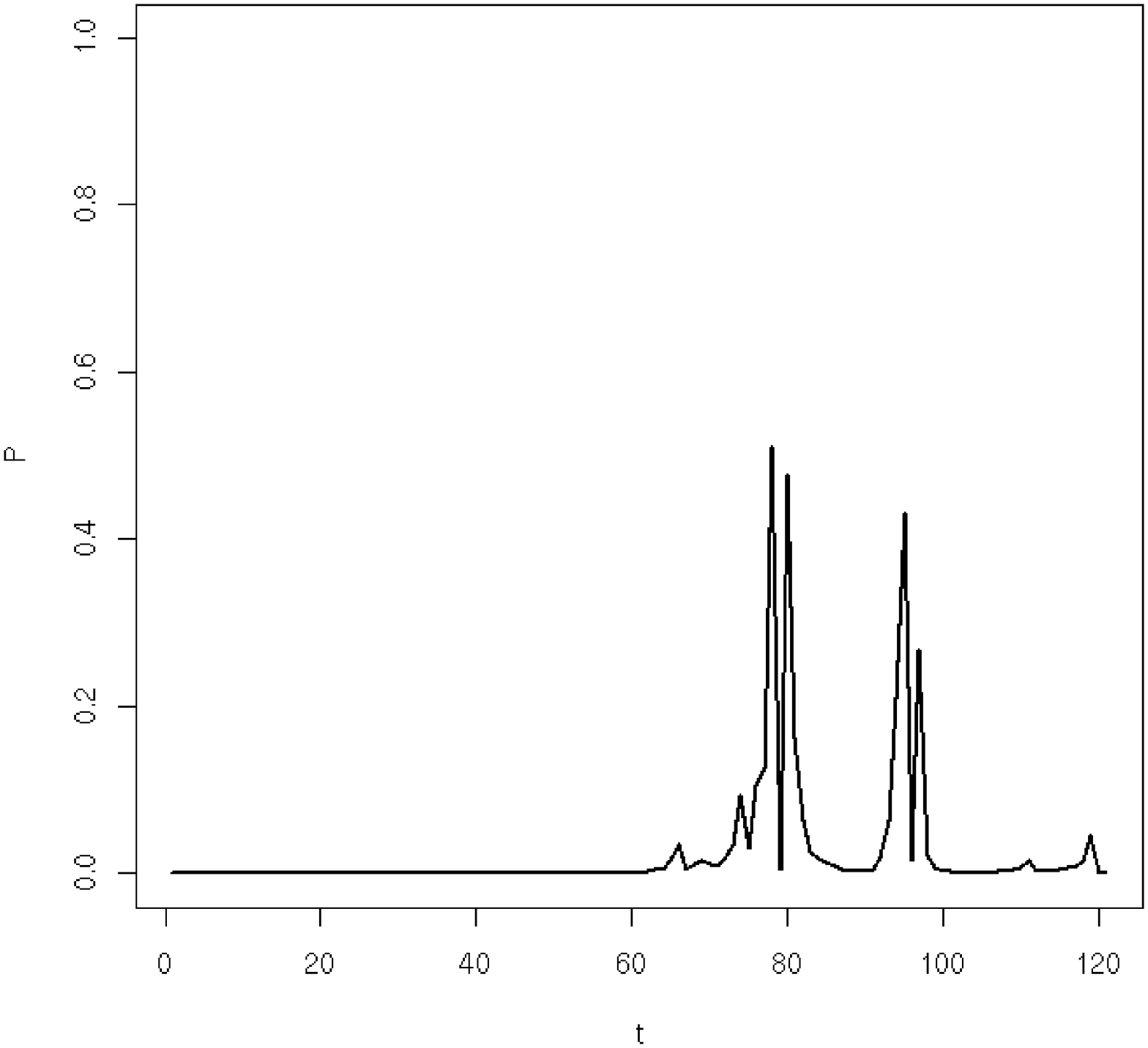}
     \end{tabular}
     \\
    $(iv)$ &
    \begin{tabular}{c}
      \includegraphics[height=4.3cm, width=6cm]{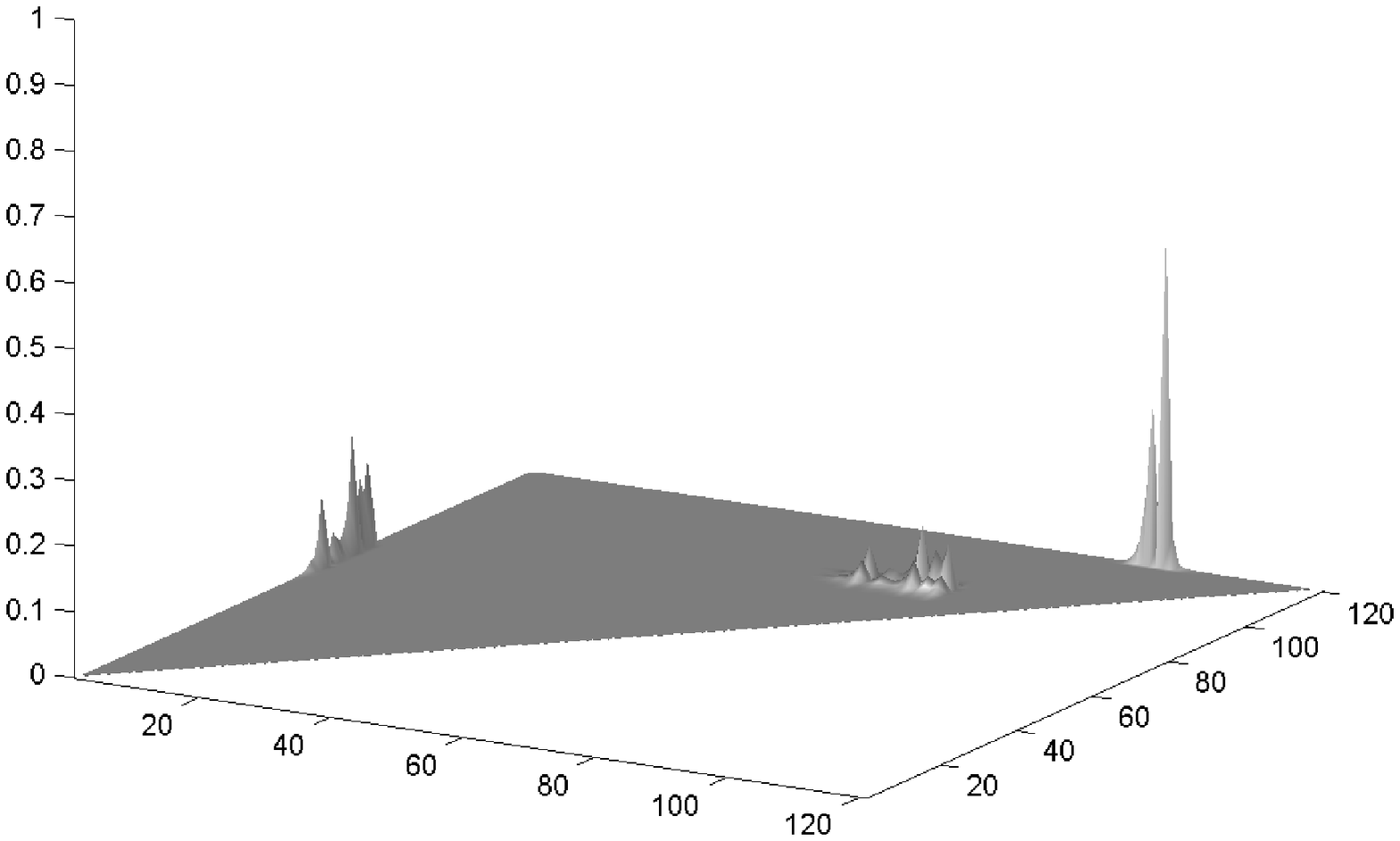}
    \end{tabular}
    &
    \begin{tabular}{c}
      \includegraphics[height=4.3cm, width=6cm]{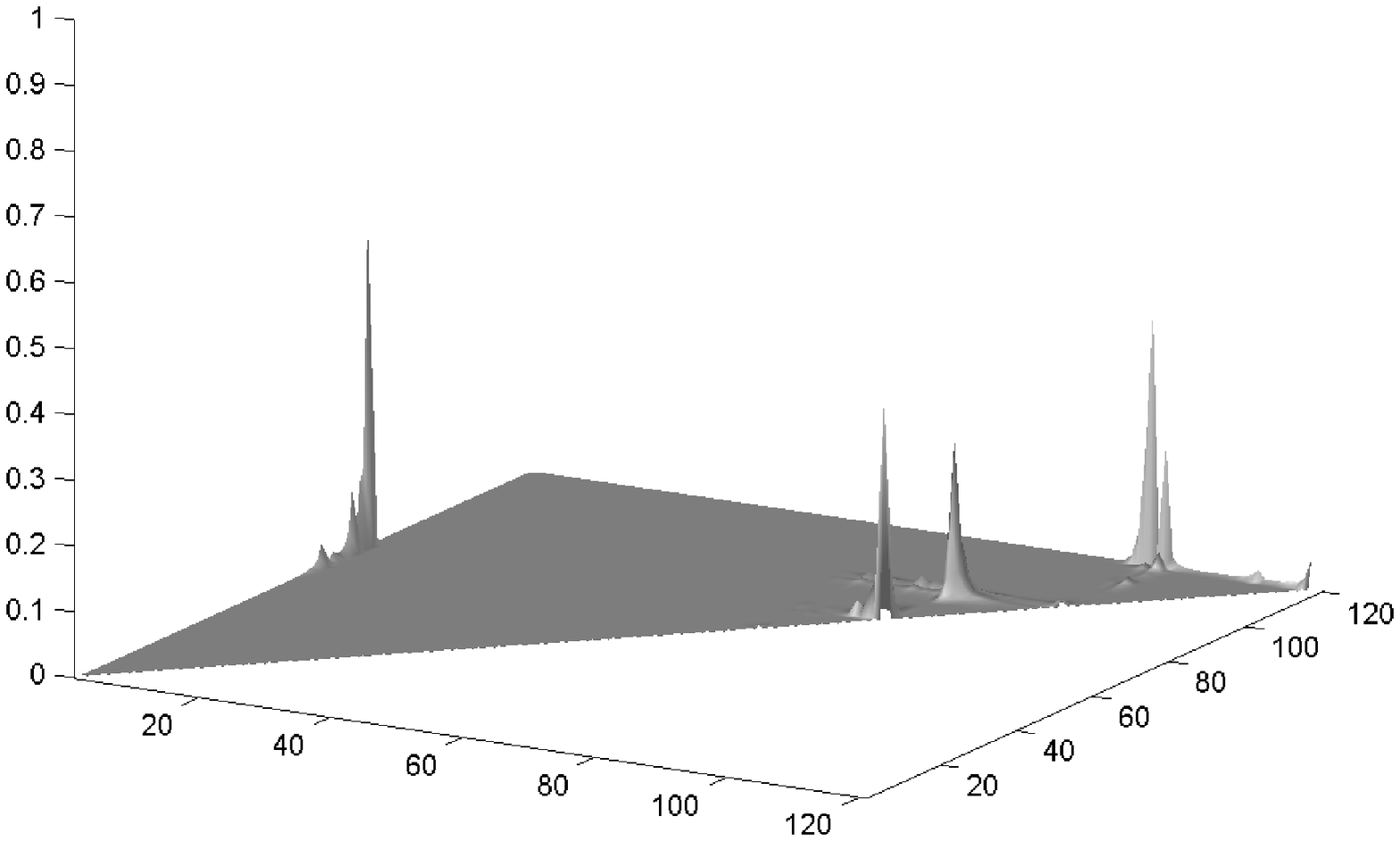}
    \end{tabular}
  \end{tabular}
  \caption{$(i)$: Best segmentation of the profile in 3 (left) and 4
    (right) segments. $\bullet$ represent the logratio as a function
    of the position along the chromosome. $-$: averaged signal of the
    segment. $\cdots$: change-point positions.  $(ii)$: Posterior
    probability that the $k$-th change-point is at position $t$
    knowing that there is either 3 (left) or 4 (right) segments.
    Probability of the first change-point: $-$, probability of the
    second change-point: $--$ and probability of the third
    change-point: $\cdots$.
     $(iii)$: Posterior probability that there is a change-point at
     position $t$ knowing that there is 3 (right) or 4 (left) segments.
    $(iv):$ 3D plot of the probability of all segments. Left panel:
    $K=3$ segments; right panel: $K=4$ segments.  $x$-axis: $t_1$,
    $y$-axis: $t_2$, $z$-axis: $S(\llbracket t_1, t_2 \llbracket)$.
    \label{Fig:ProbaPosition}}
\end{figure}

\paragraph{Model selection.}
Since the true dimension was unknown, the first issue was to choose
one. The $\ICL(K)$ criterion selected 4 segments whereas $\BIC(m)$
selected a segmentation with 3 segments (Figure
\ref{Fig:ProfilExemple}). The additional penalty term involved in ICL
does not necessarily penalise larger dimensions. In our example, ICL
selected a segmentation with a larger dimension because it was more
reliable. The choice of ICL was motivated by the relatively small gain
of entropy between dimensions 3 and 4.  This choice was also supported
by the posterior distributions of the change-points and that of the
segments shown below. The best segmentations for 3 and 4 segments are
shown on Figure \ref{Fig:ProbaPosition} ($i$).


\paragraph{Posterior probability of the change-point positions.}
The distribution of the successive change-points for dimensions 3 and
4 are shown on Figure \ref{Fig:ProbaPosition} $(ii)$.  For dimension 3, the
exact intervals with credibility $95\%$ were $\llbracket 64, 78
\rrbracket$ and $\llbracket 92, 97 \rrbracket$ for $\tau_2$ and
$\tau_3$, respectively. For dimension 4, the intervals were
$\llbracket 66, 78 \rrbracket$, $\llbracket 78, 97 \rrbracket$ and $
\llbracket 91, 112 \rrbracket$ for $\tau_2$, $\tau_3$ and $\tau_4$,
respectively.

The existence of a change-point at a given position $t$ is assessed
by posterior probability $B_K(t)$. Note that, contrarily to $B_{K,
k}(t)$, $B_K(t)$ is not a probability distribution over the
positions, because its sum is the number of change-points: $K-1$. In
our example, the posterior probabilities $B_4(t)$ presented sharper
peaks than $B_3(t)$ (see Figure \ref{Fig:ProbaPosition} $(iii)$),
which was consistent with the choice of the ICL criterion that
favours reliable segmentations.


\paragraph{Posterior probability of a segment.}
Similar conclusions were drawn from the posterior probability of the
segments. In Figure \ref{Fig:ProbaPosition} $(iv)$ each point corresponds to a segment.
A reliable dimension should display $K$ sharp peaks.
The position of the
first two segments are very uncertain for $K=3$, due to the
uncertainty of $\tau_2$. Their position were much
more certain with $K=4$. In particular, the
smallest segment from $K=4$ at positions $\llbracket 78, 79\rrbracket
$ had a relatively high probability of $0.34$.

\begin{figure}[Ht]
  \centering
  \includegraphics[height=4cm, width=5cm]{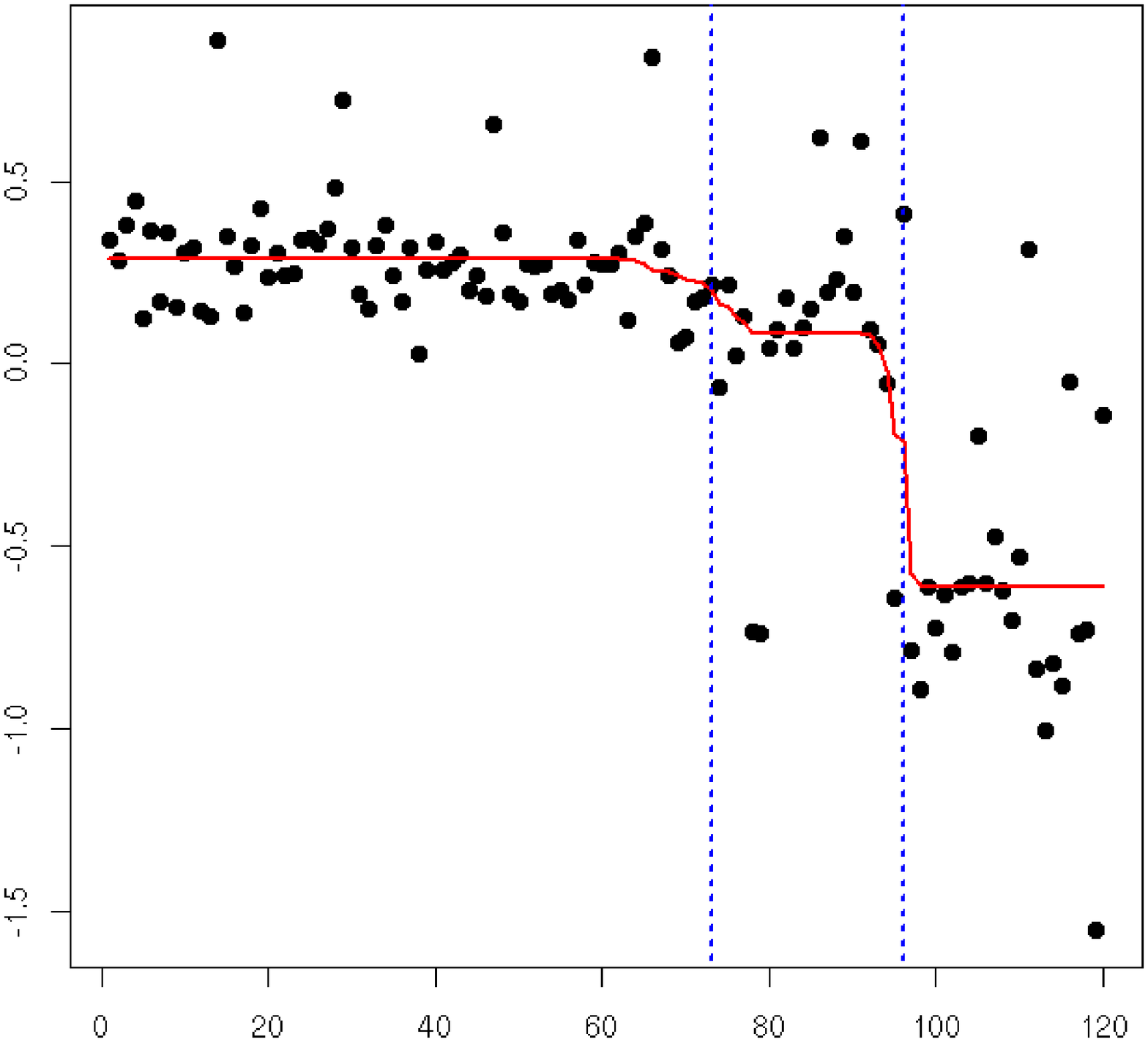}
  \includegraphics[height=4cm, width=5cm]{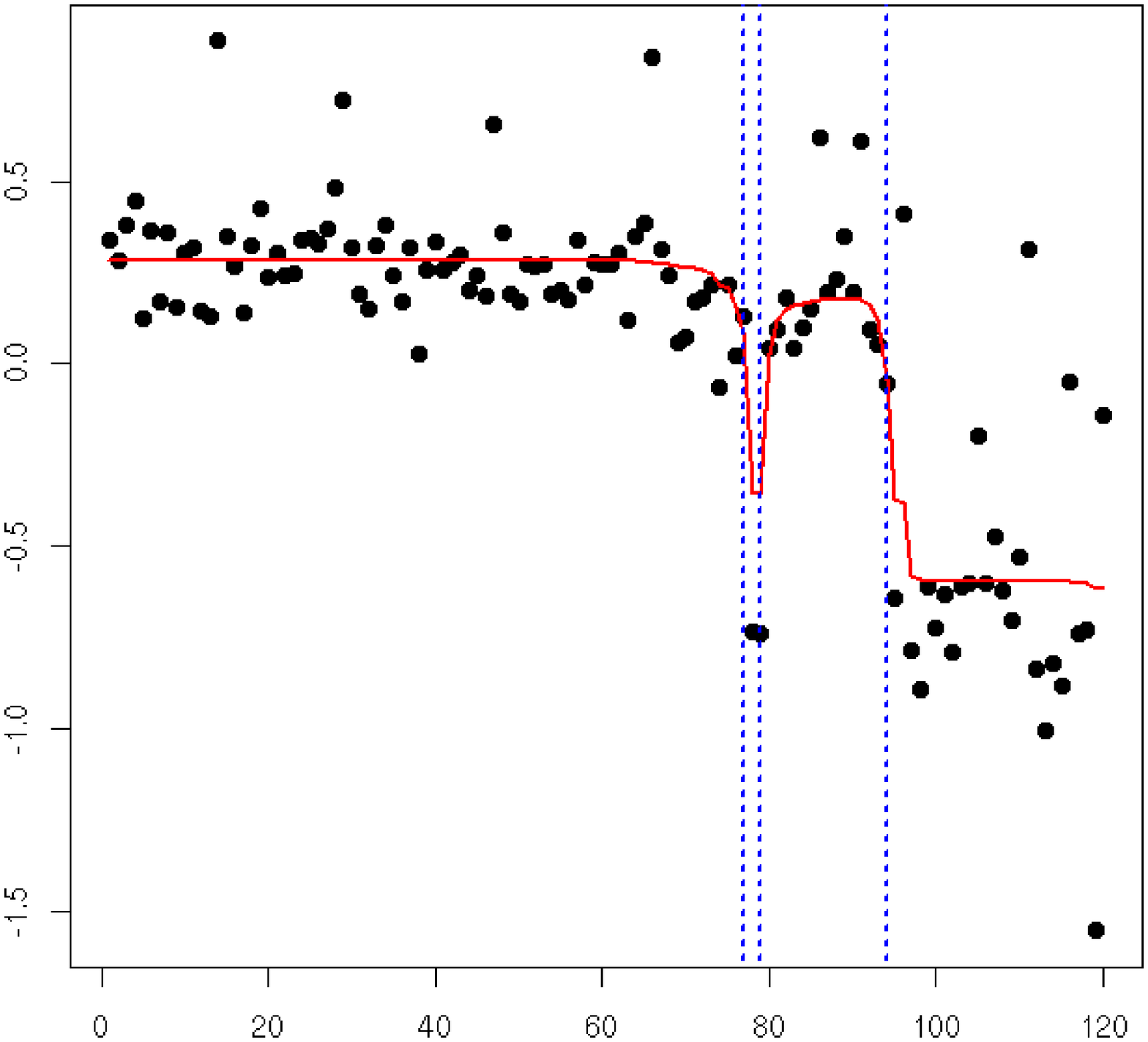}
  \includegraphics[height=4cm, width=5cm]{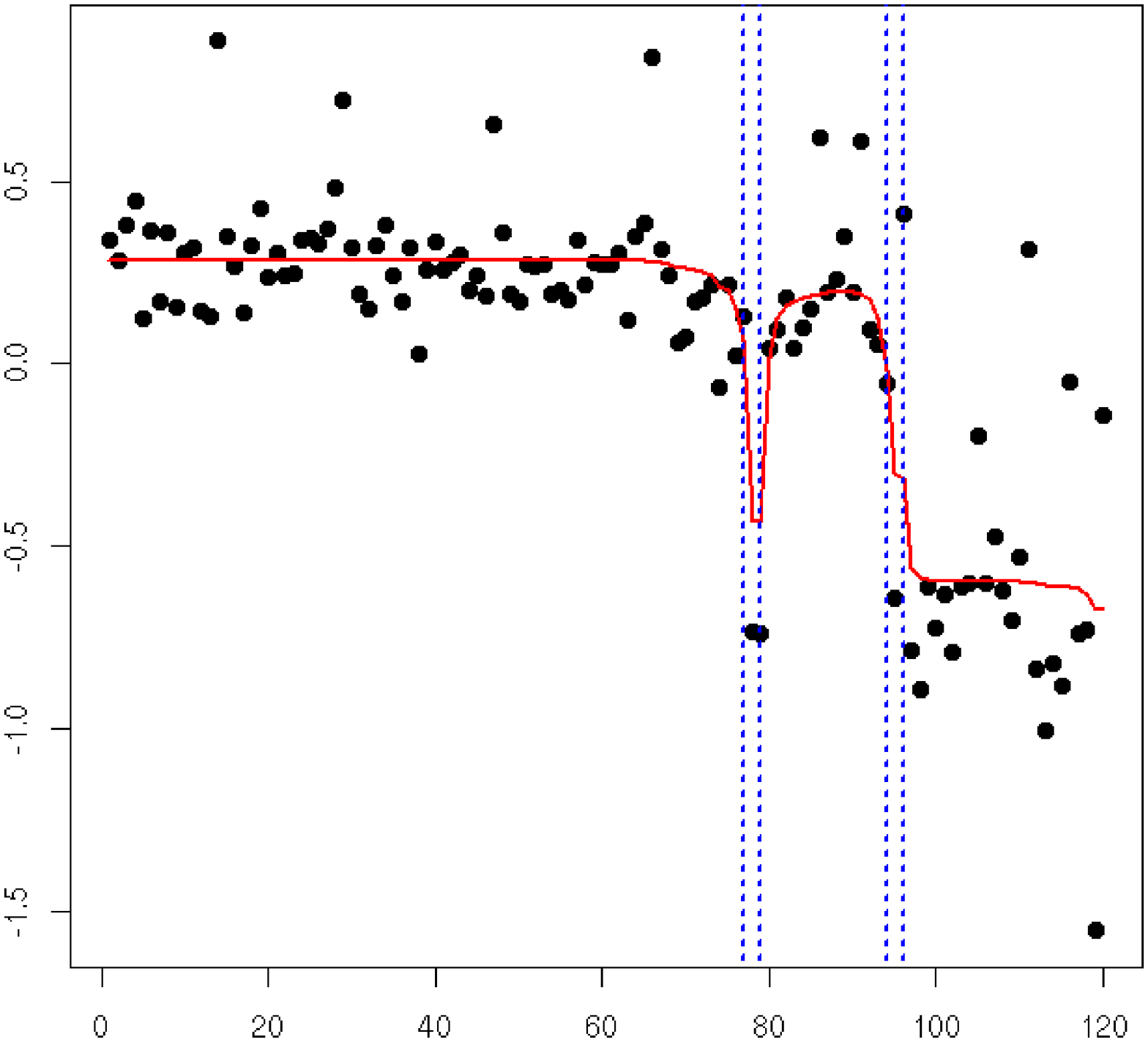}
\caption{Posterior mean of the signal; Left: $K=3$ segments;
  Center: $K=4$ segments; Right: $K=5$ segments.  $\bullet$: logratio
  as a function of the position along the chromosome. $-$: posterior
  mean of the signal.  $\cdots$: change-point positions of the best
  segmentation.
  \label{Fig:MoySeg}}
\end{figure}

\paragraph{Posterior mean of the signal.}
Similarly, the posterior mean for 3 segments was different from the
one for 4 segments (Figure \ref{Fig:MoySeg}); the former failed to capture the
small deletion at $\llbracket 78, 79\rrbracket$. As soon as $K$
exceeded 4, the posterior mean of the signal was very stable, see the
example for $K=5$ segments in Figure \ref{Fig:MoySeg}.

\paragraph{}
All presented results show that, the segmentation in 4 segments selected by the $\ICL(K)$
is more reliable than the segmentation in 3 segments selected by the
$\BIC(m)$.

\paragraph{Acknowledgements.} We thank Marie-Pierre Etienne
(AgroParisTech, UMR 518, Paris) for her helpful
advice for the writing of this paper.
We also thank Thierry Dubois (Institut Curie, dpt de Transfert) and
Emmanuel Barillot (Institut Curie, MinesParisTech, INSERM, unité U900) for their support.

\bibliography{RLR09}
\bibliographystyle{astats}

\appendix
\section{Lemma and Proofs}

\subsection{Proof of Theorem \ref{Thm:Complexity}} \label{App:ProofComplexity}

The proof of the theorem relies on the following lemma.

\begin{lem} \label{Lemma_MatrixProd}
  Let $\mathbf{A}$ be a square matrix with $n$ columns. For all $k \in
  \mathbb{N}$, we define the function $f_{\mathbf{A},k}$ as:
    \begin{eqnarray*}
    \forall (i, j) \in \llbracket 1, n \rrbracket^2  \qquad
    f_{\mathbf{A},k}(i, j) = \sum_{ (t_2 \cdots t_{k}) \ \in \ \llbracket 1, n \rrbracket^{k-1} }^{t_1= i, \ t_{k+1}=j } \ \prod_{i = 1}^{k} \mathbf{A}_{t_i, t_{i+1}}
    \end{eqnarray*}
    The $n$ elements of $\ \{ f_{\mathbf{A},k}(i, j) \}_{\{i \ \in \
      \llbracket 1, n \rrbracket \}}$ for $1 \leq k \leq K$ can all be
    computed in $O(K  n^2)$ as $$f_{\mathbf{A},k}(i, j) =
    (\mathbf{A}^k)_{i,j} $$.
\end{lem}

\paragraph{Proof of the Lemma.}
$f_{\mathbf{A},k}(i, j) = \mathbf{A}^1_{i,j} \ $ holds for $k=1$.
Suppose that $f_{\mathbf{A}}(k, i, j) = \mathbf{A}^k_{i, j }$ holds
for a given $k \in \mathbb{N}$. For $k+1$, we have:
\begin{displaymath}
  f_{\mathbf{A},k+1}(i, j) = \sum_{ (t_2 \cdots t_{k+1}) \in
  \llbracket 1, n \rrbracket^{k} }^{t_1=i, t_{k+2}=j}  \ \prod_{i = 1}^{k+1}
  \mathbf{A}_{t_i, t_{i+1}}
  =  \sum_{t = 1}^{n} \  \sum_{ (t_2 \cdots t_{k}) \in \llbracket 1, n
  \rrbracket^{k-1}
  }^{t_1=i, t_{k+1}=t} \prod_{i = 1}^{k} \mathbf{A}_{t_i, t_{i+1}} . \
  \ \mathbf{A}_{t, j}
  = \sum_{t = 1}^{n} \  f_{\mathbf{A},k}(i, t) . \ \ \mathbf{A}_{t, j}
\end{displaymath}

Using our induction hypothesis and by definition of the matrix product,
we obtain:
\begin{eqnarray*}
  f_{\mathbf{A},k+1}(i, j) & = & \sum_{t = 1}^{n} \mathbf{A}^k_{i, t}
  \mathbf{A}_{t, j} = \mathbf{A}^{k+1}_{i, j}
\end{eqnarray*}
Thus, the $K \times n$ elements of the form
$$
\left\{ f_{\mathbf{A},k}(t_1, t_{k+1}) \right\}_{\{ k \ \in \
  \llbracket 1, K \rrbracket \ \cap \ t_{k+1} \ \in \ \llbracket 1, n
  \rrbracket \} }
$$
can be computed in $O(K n^2)$ as the $t_1$-th line of matrices
$\mathbf{A}$, $\mathbf{A}^2$ $\cdots$, $\mathbf{A}^K$ respectively.
$\blacksquare$

\paragraph{Proof of the Theorem.}
For any $(t_1, ..., t_{k+1})$ in $\llbracket 1, n+1
\rrbracket^{k+1}$ such that we do not have $t_1 < t_2 \cdots <
t_{k+1}$, $\prod_{i = 1}^{k} \mathbf{A}_{t_i, t_{i+1}} = 0$.
Therefore, for all $k \in \llbracket 1, K \rrbracket $ and for all $j$
in $\llbracket 1, n \rrbracket$:
$$
\sum_{m \in \Mcal_k(\llbracket 1, j\llbracket)} F(m) = \sum_{t_1 <
  t_2 \cdots < t_{k+1}}^{t_1=1, t_{k+1}=j} \prod_{i=1}^{k}
\mathbf{A}_{t_{i},t_{i+1}} = \sum_{ (t_2, ...t_{k}) \ \in \ \llbracket
  1, n+1 \rrbracket^{k-1}}^{t_1=1, t_{k+1}=j} \prod_{i = 1}^{k}
\mathbf{A}_{t_i, t_{i+1}}
$$
Using Lemma \ref{Lemma_MatrixProd} on matrix $\mathbf{A}$
and integer $K$, we see that the $K \times (n+1)$ terms of the form
$$
\left\{ \sum_{m \in \Mcal_k(\llbracket 1, j\llbracket)} F(m)
\right\}_{k \ \in \ \llbracket 1, K \rrbracket \ \cap \ j \in
  \llbracket 1, n+1 \rrbracket }
$$
can be computed as $ \sum_{m \in \Mcal_k(\llbracket 1,
  j\llbracket)} F(m) = (\mathbf{A}^k)_{1,j} $ and that therefore they
can all be computed in $O(K n^2)$ as the first line of the
successive powers of matrix $\mathbf{A}$.

\subsection{Proof of Proposition \ref{Prop:IntQuant}} \label{App:ProofIntQuant}

\paragraph{Proof.} We first consider the posterior distribution of the
change-points. With Equation \eqref{Eq:DecompBKk}, we obtain
$$
B_{K,k}(t) = \frac{\sum_{m \in \Bcal_{K, k}(t)} P(Y | m) P(m|
  K)}{P(Y | K)} = \frac{ F_{1,t}(k-1) F_{t,n+1}(K-k+1) }{P(Y | K)}.
$$
Using Theorem \ref{Thm:Complexity}, we see that all the $F$
functions can be computed in $O(K n^2)$. $O(K^2 n)$ products and
divisions remain to be done to compute all $B_{K, k}(t)$, so the
overall complexity is in $O(K n^2)$.  The probability $B_{K}(t)$
follows straightforwardly.

We now consider the posterior distribution of the segments. We first
quote that if $t_1 = 1$, then $S_{K, 1}(1, t_2) = B_{K, 2}(t_2)$.
Similarly, when $t_2 = n+1$, we have $S_{K, K}(t_1, t_2) = B_{K, K}(t_1)$.
So we only have to consider the case where $1 < t_1 \leq t_2 < n+1$.
Since $\Scal_{K, k}(\llbracket t_1, t_2 \llbracket)$ can be decomposed as
$$
\Scal_{K, k}(\llbracket t_1, t_2 \llbracket) =
\Mcal_{k-1}(\llbracket 1, t_1 \llbracket) \times \{\llbracket t_1, t_2
\llbracket\} \times \Mcal_{K- k}(\llbracket t_2, n+1 \llbracket),
$$
we have
$$
S_{K, k}(t_1, t_2) = \frac{\sum_{m \in \Scal_{K, k}(\llbracket t_1,
    t_2 \llbracket)} P(Y | m) P(m| k)}{P(Y | K)} =
\frac{F_{1,t_1}(k-1) F_{t_1,t_2}(1) F_{t_2,n+1}(K-k)}{P(Y | K)}.
$$
Again using Theorem \ref{Thm:Complexity}, we see that all the $F$
functions can be computed in $O(K n^2)$.  We then need to compute
$O(n^2)$ products and divisions to get the $S_{K, k}(t_1, t_2)$,
thus the overall
complexity is in $O(K n^2)$. \\
The last probability comes from the definition of $S_{K}(t_1, t_2)$.
$O(K n^2)$ additions remain to be done the overall complexity is
therefore in $O(K n^2)$. $\blacksquare$

\end{document}